\documentclass[aps,prd,onecolumn,showpacs,groupedaddress,nofootinbib]{revtex4-2}
\usepackage{graphicx}
\usepackage{dcolumn}
\usepackage{bm}
\usepackage{color}
\usepackage{longtable}
\usepackage{supertabular}
\usepackage[T1]{fontenc}
\usepackage{epstopdf}
\usepackage{amsmath}
\usepackage{amssymb}
\usepackage{setspace}
\usepackage{multirow}
\usepackage{booktabs}
\usepackage[section]{placeins}
\usepackage{mathrsfs}
\usepackage{hyperref}

\usepackage{adjustbox,lipsum}

\makeatletter

\newcommand{\Rmnum}[1]{\expandafter\@slowromancap\romannumeral #1@} 
\newcommand{\bq}{\begin{equation}}
	\newcommand{\eq}{\end{equation}}
\newcommand{\bqn}{\begin{eqnarray}}
	\newcommand{\eqn}{\end{eqnarray}}
\newcommand{\nb}{\nonumber}
\newcommand{\lb}{\label}

\makeatother

\begin{document}
	
	\title{Evolution of black hole echo modes and the causality dilemma}
	
	\author{Ramin G. Daghigh\textsuperscript{1}}\email[E-mail: ]{ramin.daghigh@metrostate.edu}
	\author{Guan-Ru Li\textsuperscript{2}}\email[E-mail: ]{guanru.li@unesp.br}
	\author{Wei-Liang Qian\textsuperscript{3,2,4}}\email[E-mail: ]{wlqian@usp.br}
	\author{Stefan J. Randow\textsuperscript{5}}\email[E-mail: ]{stefan.randow@my.metrostate.edu}
	
	\affiliation{$^{1}$ Natural Sciences Department, Metropolitan State University, Saint Paul, Minnesota, USA 55106}
	\affiliation{$^{2}$ Faculdade de Engenharia de Guaratinguetá, Universidade Estadual Paulista, 12516-410, Guaratinguetá, SP, Brazil}
	\affiliation{$^{3}$ Escola de Engenharia de Lorena, Universidade de São Paulo, 12602-810, Lorena, SP, Brazil}
	\affiliation{$^{4}$ Center for Gravitation and Cosmology, College of Physical Science and Technology, Yangzhou University, Yangzhou 225009, China}
	\affiliation{$^{5}$ Mathematics and Statistics Department, Metropolitan State University, Saint Paul, Minnesota, USA 55106}

	\begin{abstract}
		It has been shown that black hole quasinormal modes are subject to spectral instability, typically triggered by metric perturbations.
		These perturbations, which can introduce a minor bump in the effective potential of the wave equation, give rise to a novel branch of asymptotic quasinormal modes, dubbed the {\it echo modes}, which lie mainly parallel to the real frequency axis.
		This study explores the evolution of the echo modes and their interplay with the outward spiral motion observed in low-lying quasinormal modes.
		As the bump in the effective potential moves away from the central black hole, the echo modes collectively shift toward the real axis, with the spacing between successive modes decreasing uniformly.
		This collective motion occurs simultaneously with the spiral of the low-lying modes until the echo modes eventually take over the fundamental quasinormal mode.
		In the time domain, such a takeover coincides with a transition point for the temporal waveform, where the distinction between the original black hole's ringdown and the echoes becomes clear.
		This marks a transition in the characteristics of the waveform from primarily damped oscillations, dominated by the damping rate of the fundamental mode, to echo waves, characterized by periodic echo pulses.
		We argue that this phenomenon is universal by employing analytical and numerical analyses.
		We first elucidate our arguments using explicit but simplified toy models, where the effective potential barriers are disjoint.
		The derivations are then generalized to scenarios where perturbations are introduced on top of a black hole metric with a continuous effective potential.
		The observational implications, particularly the causality dilemma, are elaborated.
		We show that the echo modes can be extracted by applying the Fourier transform to ringdown waveforms, which can be important for gravitational wave observations.  
	\end{abstract}
	
	\date{Jan. 7th, 2025}
	
	\maketitle
	
	\newpage
	
	\section{Introduction}\label{sec1}
	
	The extreme gravity in the vicinity of black holes is one of the most captivating concepts in theoretical physics.
	The successful detection of gravitational waves from binary black hole mergers by the LIGO and Virgo collaborations~\cite{agr-LIGO-01, agr-LIGO-02, agr-LIGO-03, agr-LIGO-04} has inaugurated a new era for observational astrophysics.
	This achievement has further inspired many ongoing spaceborne projects, such as LISA~\cite{agr-LISA-01}, TianQin~\cite{agr-TianQin-01, agr-TianQin-Taiji-review-01}, and Taiji~\cite{agr-Taiji-01}, where direct observation of ringdown waveforms has been speculated to be plausible~\cite{LISA-ringdown}.
	
	Quasinormal modes (QNMs) are the natural vibrational modes of black holes that determine the shape of the observed ringdown waveform.
	The significance and stability of QNMs, when discontinuities are introduced to the Regge-Wheeler potential, were initially investigated in~\cite{Nollert1, Nollert2}.
	These investigations were refined and expanded upon in~\cite{DGM-significance, Qian-QNM-Significance}, followed immediately by multiple papers on the stability of the QNM spectrum such as~\cite{QNM-Significance1, QNM-Significance2, QNM-Significance3, QNM-Significance4}.
	There is now strong evidence indicating that the asymptotic QNMs are mainly unstable against metric perturbations~\cite{QNM-Significance1}, particularly when discontinuities are introduced to the effective potential of the wave equation in linearized gravity.
	
	Spectral instability is not exclusive to high overtones.
	The authors of~\cite{Cheung} investigated the instability of the fundamental QNM by introducing a perturbative bump to the effective potential.
	As the bump moves away from the central black hole, the fundamental mode initially spirals away from its original value and later moves, accompanied by other modes in the spectrum, towards the real frequency axis on the complex plane.
	Recently, we showed that the universality of the initial spiral occurring for low-lying modes can be understood from an analytic perspective~\cite{Qian:2024iaq}.
	The proof was given for both truncated effective potentials and those constituted by disjoint potential barriers.
	
	Notably, a paradox is related to the QNM instability initially pointed out by Nollert~\cite{Nollert1}.
	On the one hand, as pointed out in~\cite{Cheung}, the modification to the fundamental mode grows with the distance of the perturbative bump from the central black hole.
	Therefore, the effect is expected to be significant when the perturbative bump is located at a considerable distance from the black hole.
	On the other hand, an insignificant perturbation planted far away from the physical system of interest is not supposed to affect the observational outcome.
	Based on numerical calculations of the time-domain waveforms, the obtained ringdown profiles remain largely intact in the presence of small discontinuities~\cite{Nollert1, DGM-significance, Qian-QNM-Significance, EnvEffect, Ringdown-Stability, BHSpectroscopy}.
	Although, intuitively, the instability of the QNM spectrum might imply a significant challenge to black hole spectroscopy~\cite{BHSpectroscopy}, the time-domain stability of the ringdown waveform seems to indicate otherwise.
	These results call for a detailed analysis of the feasibility of black hole spectroscopy in the context of spectral instability, which is still lacking in the literature.
	
	Interestingly, the above arguments are also in line with a causality dilemma.
	To elucidate the dilemma, let us explicitly consider the scenario where the bump is disjoint from the effective potential and is placed far away from the black hole horizon.
	For a pure black hole, the asymptotic QNMs lie parallel to the imaginary axis, while for the perturbed metric, the spectrum is significantly deformed and lines up along the real axis.
	In other words, the frequency domain Green's function drastically differs between the two cases.
	Let us now assume an observer sits at a finite distance from the black hole and detects the emitted gravitational wave signals.
	Intuitively, one might expect the observer to extract the QNMs from the measured ringdown waveforms successfully and conclude that the asymptotic QNMs are primarily lined up along the real axis.
	However, this raises the question of how the measured waveform would ever carry information about the perturbative bump before the initial data can interact with the bump causally.
	Such a causality concern was discussed by Hui~\textit{et al.}~\cite{Hui}, where the authors tackle the problem elegantly by explicitly evaluating the time-domain waveform from the frequency-domain Green's function.
	It was shown that the Green's function of the perturbed metric can be written as a summation of a geometric series where the first term is identical to the Green's function of the original black hole.
	At an earlier instant, specifically, before the initial data could pick up any information on the bump permitted by causality, the time-domain waveform will only receive non-vanishing contributions from the first term of Green's function.
	Consequently, the resulting waveform is precisely the same as the original black hole.
	As time increases, the inverse Laplace transform picks up more and more terms in the series due to Jordan's lemma, and the waveform becomes gradually deformed, reflecting more contributions from the metric perturbation.
	This provides a \textit{time-dependent} picture of how the waveform derives contributions from the underlying Green's function.
	Due to the discrete and sequential nature of the formalism, echoes naturally merge, with their period determined by the ratio of consecutive terms in the geometric series and the number of repetitions dictated by the effective inclusion of terms from the summation.
	
	The study in \cite{Hui} demonstrates causal consistency in scenarios involving disjoint bounded potentials, specifically elucidating why the initial waveform matches that of an unperturbed black hole.
	However, it still leaves a few unattended questions, particularly regarding the whereabouts of the poles of the Green's function (also known as QNMs).
	First, the poles arising from all individual terms of the geometric series constituting the s-domain Green's function are fixed and identical to those of the original black hole, regardless of the bump's location.  
	In contrast to fixed poles, the fundamental mode was observed~\cite{Cheung, Yang:2024vor, Ianniccari:2024ysv, Qian:2024iaq} to be unstable and spirals outward from its original location.
	Also, the observed asymptotic QNMs lying parallel to the real axis, dubbed the {\it echo modes}, do not seem to hold a place in the mathematical formalism proposed in~\cite{Hui}. 
	This makes it difficult, if not impossible, to construct the echoes that appear after the ringdown waveform.
	Moreover, as elaborated below, the echo mode will never appear if one only considers a finite number of terms in the summation.
	In this regard, the causality dilemma is elevated at the cost of removing the migration of the fundamental mode and the presence of echo modes.
	
	Is it possible to elaborate on a unified picture that consistently embraces both arguments?
	It is understood that for disjoint potentials and given the bump's position, we can precisely extract information on the black hole by only measuring the waveform up to the instant related to the emergence of the first echo pulse.
	However, for a more realistic scenario, the black hole metric and the perturbations are continuous rather than disjoint.
	In other words, the information on the original black hole's QNM and its significantly distorted spectrum are intrinsically entangled, while the echo waveform may not be easily distinguished from the black hole's quasinormal oscillations. 
	For such a scenario, is it still feasible to extract pertinent information from the observational data?
	Can we generalize the existing derivations, and does the physical picture remain unchanged?
	
	The present study is motivated by the above considerations.
	We further explore the properties of the echo modes and aim to answer the above questions partially.
	By comparing with the picture proposed in~\cite{Liu:2021aqh, Qian:2024iaq}, we argue that both the emergence of echo modes and the deviation in the black hole's fundamental mode should be understood as a collective effect from an infinite number of terms constituting the geometric series provided in \cite{Hui}.
	Through numerical and analytical calculations, we demonstrate that the echo modes evolve as the bump's location moves away from the black hole horizon.
	We point out an intriguing interplay between the evolution of the echo modes and the spiral of the fundamental mode.
	Specifically, echo modes eventually take over the spiraling fundamental mode, leading to the phenomenon observed in~\cite{Cheung}. 
	One of the echo modes stands out and replaces the fundamental mode, giving rise to a jump in the real part of the quasinormal frequency.
	However, since the imaginary part (damping rate) of the echo modes is similar in magnitude, they always collectively contribute to the waveform, giving rise to echo pulses.
	We claim the universality of the above picture by generalizing the derivation to the cases where the metric perturbation is placed onto a continuous black hole metric.
	
	The remainder of the paper is organized as follows.
	In Sec.~\ref{sec2}, we give a detailed account of the spectrum constituted by the echo modes and their evolution.
	We start by focusing on a class of effective potentials where the potential barriers are disjoint.
	After exploring a few particular cases, including the double delta-function and double square barriers, we give a general analytic argument for the universality of how the QNM spectrum evolves for such effective potentials.
	In Sec.~\ref{sec3}, we elaborate on an interplay between the spiral of the fundamental mode and the evolving echo modes.
	The overtaking process in the frequency domain is illustrated numerically.
	In Sec.~\ref{sec4}, we generalize our results to the scenario where the metric perturbations are introduced to black hole effective potentials that initially cover the entire spatial coordinate.
	The properties of the corresponding time-domain waveform are investigated in Sec.~\ref{sec5}.
	Further illustrations of the feasibility of extracting echo modes are given. 
	We reflect on the dilemma about causality and the location of the poles of Green's function.
	The last section is dedicated to the concluding remarks.
	
	\section{Universality in emergence and evolution of the echo modes in disjoint potential barriers}\label{sec2}
	
	In this section, we explore the properties of echo modes and their evolution from an analytic perspective.
	Before tackling a generalized scenario, we elaborate on our arguments using explicit but simplified toy models, where the effective potential barriers are defined on disjoint domains.
	
	In this work, following \cite{Hui}, we model the perturbations to the black hole metric using the superposition of two disjoint bounded potentials, $V_1(x)$ and $V_2(x)$.   
	We denote $V_1(x)$ as the effective potential of the original black hole and $V_2(x)$ as the perturbative bump.
	The resulting master equation for black hole QNMs takes the following form
	\begin{equation}
		\left[\partial_t^2-\partial_x^2 +V(x)\right]\Psi(t,x)=0 ,
		\label{Eq: waveEQ}
	\end{equation}
	where
	\bqn
	V(x)=V_1(x)+V_2(x-L) ,
	\eqn  
	in which $L$ measures the separation between the two potentials, and the magnitude of $V_2(x)$ is understood to be much smaller than that of $V_1(x)$.

	Subsequently, the Wronskian of the frequency-domain retarded Green's function has the form~\cite{Hui}
	\begin{equation}
		W_{1+2}(\phi_-,\phi_+)= -2 i \omega \left[ m_1^{++}(\omega) m_2^{++}(\omega)+ e^{2 i \omega L} m_1^{-+}(\omega)m_2^{+-}(\omega)\right] ,
		\label{eq: Wronskian}
	\end{equation}
	where the ingoing and outgoing waves are
	\bqn
	\phi_+(\omega,x) &=& m_1^{++}e^{-i\omega x}+m_1^{-+}e^{i\omega x},\lb{inW} \\
	\phi_-(\omega,x) &=& -m_2^{+-}e^{i\omega L}e^{-i\omega x}+m_2^{-+}e^{-i\omega L}e^{i\omega x},\lb{outW}
	\eqn
	and $m_{1(2)}^{\pm\pm}$ are elements of the transit matrix
	\bqn
	M_{1(2)} =  \begin{pmatrix}m_{1(2)}^{++}&m_{1(2)}^{+-}\\m_{1(2)}^{-+}&m_{1(2)}^{--}\end{pmatrix} 
	\eqn
	defined as transmission and reflection coefficients for a left(right)-going incident wave (denoted by the superscript $\pm$) coming from the left of the potential barrier $V_{1(2)}$.
	
	These quantities are functions of the frequency 
	\bqn
	\omega = \omega_R + i\omega_I .
	\eqn 
	The QNMs are governed by Wronskian's roots, namely,
	\begin{equation}
		m_1^{++}(\omega) m_2^{++}(\omega)+ e^{2 i \omega L} m_1^{-+}(\omega)m_2^{+-}(\omega)=0 . \label{eq: QNMs}
	\end{equation}
	Note that the roots of $m_1^{++}(\omega)$ are the QNMs of the potential $V_1(x)$, which correspond to the unperturbed black hole metric.
	For later convenience, we note that Eq.~\eqref{eq: QNMs} can be rewritten into the form
	\bqn
	\mathbf{f}(\omega) = -e^{2i\omega L} ,\label{theEq2Solve}
	\label{Eq: fw}
	\eqn
	where we have defined
	\begin{equation}
		\mathbf{f}(\omega)= \frac{m_1^{++}(\omega)m_2^{++}(\omega)}{m_1^{-+}(\omega)m_2^{+-}(\omega)} .
		\label{eq: fw}
	\end{equation}
	
	We elaborate on a few specific examples before giving arguments from a more general perspective.
	For the first example, we consider a double delta-function potential barrier.
	Specifically, we have 
	\bqn
	V_1(x) &=& v_1 \delta(x) , \nb\\
	V_2(x) &=& v_2 \delta(x-L) .\lb{twoDelta}
	\eqn
	It is not difficult to show~\cite{Hui} that the Wronskian possesses the form
	\begin{equation}
		W_{1+2}(\omega)=\frac{1}{-2 i \omega} \left[ (-2 i \omega+v_1)(-2 i \omega+v_2)-v_1 v_2  e^{2 i \omega L}\right].
		\label{W12simplified}
	\end{equation}
	By comparing against the form~\eqref{eq: fw}, we have 
	\begin{equation}
		\mathbf{f}(\omega)=- \frac{(-2 i \omega+v_1)(-2 i \omega+v_2)}{v_1v_2 }.
		\label{fOmegaEx1}
	\end{equation}
	
	We assume the perturbation is insignificant, namely, $v_2 \ll  v_1$. 
	Also, we are interested in asymptotic modes for which $\omega_R \gg |\omega_I|$.  
	Subsequently, one can approximately rewrite Eq.~\eqref{fOmegaEx1} as
	\begin{equation}
		\mathbf{f}(\omega)\approx 
		\frac{4}{v_1 v_2}\omega_R^2\left[1 +\frac{i (v_1-4\omega_I)}{2 \omega_R}\right] ~,
		\label{fOmegaApp1}
	\end{equation}
	where an expansion has been carried out in terms of $1/\omega_R$.
	At the lowest order, one can ignore $\omega_I$ and only consider the most dominant contribution from $\omega_R$. 
	In this case, the r.h.s. of Eq.~\eqref{fOmegaApp1} is mostly a real number, and the imaginary part of Eq.~\eqref{theEq2Solve} gives
	\begin{equation}
		\omega_R \approx \left(n+\frac12\right)\frac{\pi}{L} ,
		\label{Eq: DDeltawR}
	\end{equation}
	where $n\gg 1$ is the overtone index and the factor $e^{i\pi}=-1$ on the r.h.s.\ of Eq.~\eqref{theEq2Solve} leads to an additional shift of $\frac12$ to the overtone number $n$.
	To estimate the dominant contribution from $\omega_I$, we note that the correction on the exponential on the r.h.s.\ of Eq.~\eqref{theEq2Solve} plays a more significant role than its counterpart on the r.h.s. of Eq.~\eqref{fOmegaApp1}.
	Specifically, one retains the first-order term of the r.h.s. of Eq.~\eqref{fOmegaApp1}, namely,
	\begin{equation}
		\mathbf{f}(\omega) \approx 
		\frac{4\omega_R^2}{v_1 v_2} ~ ,
		\label{FformOmega}
	\end{equation} 
	and considers the real part of the equality in Eq.~\eqref{theEq2Solve} that gives
	\begin{equation}
		\omega_I\approx -\frac{1}{2L}\ln \frac{4\omega_R^2}{v_1 v_2}
		\approx -\frac{1}{2L}\left[2\ln \left(n+\frac12\right)+\ln\frac{4\pi^2}{v_1v_2}-2\ln L \right].
		\label{Eq: DDeltawI}
	\end{equation}
	This is the dominant behavior in the QNM spectrum of a double delta-function barrier.
	The real part of the frequency is more significant than the imaginary part for asymptotic modes, where $n\gg 1$.
	Although the imaginary part of the frequency increases as $n$ increases, the QNM spectrum governed by Eqs.~\eqref{Eq: DDeltawR} and~\eqref{Eq: DDeltawI} line primarily parallel to the real axis.
	The distance between successive modes along the real axis is $\frac{\pi}{L}$, governed by the distance between the locations of the two delta functions.
	As $L$ increases, the distance between successive modes and the distance to the real axis decrease since both quantities are inversely proportional to $L$.

	One can proceed further to a higher order.
	Specifically, one substitutes Eq.~\eqref{Eq: DDeltawI} into the r.h.s. of Eq.~\eqref{fOmegaApp1} and reads off the modulus and the phase of $\mathbf{f}(\omega)$.
	By comparing the real and imaginary parts of the equality, one finds
	\begin{equation}
		\omega_R \approx \left(n+\frac{1}{2}\right)\frac{\pi}{L} +\frac{1}{2L}\arctan\frac{ (v_1-4\omega_I)L}{2 \tilde{n}\pi}
		\approx  \left(n+\frac{1}{2}\right)\frac{\pi}{L} +\frac{ 4\ln \tilde{n}+2\ln\frac{4\pi^2}{v_1v_2 L^2} +v_1 L}{4 \tilde{n}\pi L}
		\label{Eq: DDeltawR2}
	\end{equation}
	and
	\begin{equation}
		\omega_I \approx -\frac{1}{2L}\left[2\ln \left(n+\frac{1}{2}\right)+\ln\frac{4\pi^2}{v_1v_2}-2\ln L +\frac{ 4\ln \tilde{n}+2\ln\frac{4\pi^2}{v_1v_2 L^2} +v_1 L}{2 \tilde{n}^2\pi^2 }
		\right],
		\label{Eq: DDeltawI2}
	\end{equation}
	where $\tilde{n}=n+1/2$.
	Eqs.~\eqref{Eq: DDeltawR2} and~\eqref{Eq: DDeltawI2} do not alter the main feature observed in Eqs.~\eqref{Eq: DDeltawR} and~\eqref{Eq: DDeltawI}.
	
	As the second example of disjoint effective potentials, we consider a double square barrier, which was considered in~\cite{Hui, Cheung}.
	It consists of two square barriers of different heights.
	The more significant potential barrier of a height $h$ represents the effective potential of the black hole, while the minor barrier with a height $\epsilon$ corresponds to the perturbation.
	The explicit form of the potential is 
	\begin{equation}
		V(x) = \left\{ \begin{array}{ll}
			0 & x \le -\frac{\sigma_0}{2}~\\  \\
			h & -\frac{\sigma_0}{2} < x < \frac{\sigma_0}{2}~\\    \\
			0  & \frac{\sigma_0}{2} \le x \le L-\frac{\sigma}{2}~\\    \\
			\epsilon & L-\frac{\sigma}{2} < x < L+\frac{\sigma}{2}~\\    \\
			0 & x \ge L+\frac{\sigma}{2}
		\end{array}
		\right. ~,       
		\label{Vsquare}
	\end{equation}
	where $\epsilon \ll h$.
	For this potential, one finds
	\begin{eqnarray}
		m_1^{++}(\omega)&=&e^{i\sigma_0 \omega}\left[e^{i\sigma_0 \bar{\omega}}(\omega-\bar{\omega})^2-e^{-i\sigma_0 \bar{\omega}}(\omega+\bar{\omega})^2\right] ,\nonumber  \\
		m_2^{++}(\omega)&=&e^{i\sigma \omega}\left[e^{i\sigma \hat{\omega}}(\omega-\hat{\omega})^2-e^{-i\sigma \hat{\omega}}(\omega+\hat{\omega})^2\right] ,\nonumber  \\
		m_1^{-+}(\omega)&=& 2 h \sin{(\sigma_0 \bar{\omega})} ,\nonumber \\
		m_2^{+-}(\omega)&=& 2 \epsilon \sin{(\sigma \hat{\omega})}  ,
		\label{eq: TransitMatrix2S}
	\end{eqnarray}
	where $\bar{\omega}=\sqrt{\omega^2-h}$ and $\hat{\omega}=\sqrt{\omega^2-\epsilon}$. 
	As one might have expected for a perturbative bump, $m_2^{+-}(\omega)\to 0$  as $ \epsilon\to 0$.

	We now can estimate the asymptotic QNMs by assuming $\omega_R \gg |\omega_I|$ and $\omega_R^2 \gg  h \gg \epsilon$.  With these assumptions, Eq.~\eqref{eq: fw} gives
	\begin{equation}
		\mathbf{f}(\omega)\approx 
		\frac{4 \omega_R^4}{h \epsilon \sin[\sigma_0 (\omega-h/\omega_R)]\sin[\sigma (\omega-\epsilon/\omega_R)]}   e^{i\frac{\sigma_0 h+\sigma \epsilon}{2\omega_R}} \left( 1-4 i \frac{\omega_I}{\omega_R} -\frac{h+\epsilon}{2\omega_R^2}\right) ~.
		\label{f2square}
	\end{equation}
	The modulus has the form 
	\begin{equation}
		f(\omega)\equiv |\mathbf{f}(\omega)|\approx 
		\frac{8 \omega_R^4}{h \epsilon \sqrt{\cosh(2\sigma_0 \omega_I)-\cos[2\sigma_0(\omega_R-h/\omega_R)]}\sqrt{\cosh(2\sigma \omega_I)-\cos[2\sigma(\omega_R-\epsilon/\omega_R)]}} \left( 1-\frac{h+\epsilon}{2 \omega_R^2}\right) ~.
		\label{ma112}
	\end{equation}
	Employing a similar strategy, we have 
	\bqn
	\omega_R \approx \left(n+\frac12\right)\frac{\pi}{L} ,\lb{omegaR2squre}
	\eqn
	and
	\begin{equation}
		\omega_I  \approx -\frac{1}{2L}\left[ \ln \frac{8 \omega_R^4}{h \epsilon }-\frac12\ln\left[\cosh(2\sigma_0 \omega_I)-\cos(2\sigma_0 \omega_R)\right]-\frac12\ln\left[\cosh(2\sigma \omega_I)-\cos(2\sigma \omega_R)\right]     \right].
		\label{omegaI2squre}
	\end{equation}
	To the lowest order, one ignores the contribution from $\omega_I$, and Eq.~\eqref{omegaI2squre} simplifies to 
	\begin{equation}
		\omega_I  \approx -\frac{1}{2L}\ln\left[\frac{8 \omega_R^4}{h \epsilon } \right]
		\approx -\frac{2}{L}\ln\omega_R \approx -\frac{2}{L}\ln \left(n+\frac12\right)
		\label{omegaI2squreA}
	\end{equation}
	consistent with the results obtained for the double delta-function potential.
	
	It is not difficult to show that the features for asymptotic QNMs drawn from the above particular examples are universal in the context of two disjoint bounded potentials. 
	Let us rewrite $\mathbf{f} (\omega)$ in terms of its modulus and argument, namely,
	\bqn
	\mathbf{f}(\omega) = f(\omega) e^{i \phi(\omega)}.
	\eqn
	By equating the real and imaginary parts of Eq.~\eqref{theEq2Solve}, we have
	\begin{equation}
		\omega_R =\left(n+\frac{1}{2}\right)\frac{\pi}{L} +\frac{\phi(\omega)}{2L},
		\label{Eq: wR}
	\end{equation}
	as $e^{2 i \omega_R L}=-e^{i \phi(\omega)}$ and
	\begin{equation}
		\omega_I=-\frac{\ln {f}(\omega)}{2L} .
		\label{Eq: wI}
	\end{equation}
	It is noted that the above results are largely impractical as they are implicit.
	In the lowest order, for asymptotic modes $\omega_R \gg |\omega_I|$, the contribution of the phase $\phi$ is negligible.
	Specifically, we have
	\bqn
	\omega_R \approx \left(n+\frac12\right)\frac{\pi}{L} ,
	\label{Eq: wRGen}
	\eqn
	and 
	\bqn
	\omega_I \approx -\frac{\ln {f}(\omega_R)}{2L} = -\frac{\ln {f}\left[\left(n+\frac12\right)\frac{\pi}{L}\right]}{2L} .
	\label{Eq: wIGen}
	\eqn
	One may further refine the above results by including higher-order contributions.
	As shown explicitly in the above examples, this can be achieved by considering the specific form of $\mathbf{f}(\omega)$, expanding it in terms of $\frac{\omega_I}{\omega_R}$, and substituting the obtained expressions back into the derivations of Eqs.~\eqref{Eq: wRGen} and~\eqref{Eq: wIGen}. 
	Nonetheless, we have seen that the analysis for the lowest order suffices to demonstrate the main features of the asymptotic QNMs.
	As discussed above and pointed out in~\cite{Liu:2021aqh}, these asymptotic modes appear due to the perturbative bump and are responsible for the echoes observed in the waveforms.
	For a frequency-domain Green's function with poles lined up parallel to the real axis, the inverse Fourier transform results in a time-domain waveform with periodic modulation enveloping the quasinormal oscillations.
	
	The physical interpretation of the novel spectrum governed by Eqs.~\eqref{Eq: wRGen} and~\eqref{Eq: wIGen} is straightforward.
	It corresponds to a quasi-bound state between the effective potential and the perturbative bump, which can be interpreted as a finite potential well.
	The real part, Eq.~\eqref{Eq: wRGen}, is strongly reminiscent of the energy of the stationary state of the corresponding Schr\"odinger equation~\eqref{Eq: waveEQ} if this potential well were replaced by an infinite square well.
	On the other hand, the imaginary part, Eq.~\eqref{Eq: wIGen}, reflects energy leakage due to the finiteness of the potential wall.
	The latter is dictated by the transition coefficients of the two potential walls and the width of the potential well that measure the energy flux and the total energy stored in the potential well, respectively.
	The observation that the magnitude of $\omega_I$ increases with increasing overtone number $n$ reflects the fact that the ratio between the transmission and incident amplitudes increases monotonically with the energy.
	Nevertheless, such a spectrum appears even when the black hole potential and the perturbation do not explicitly form a potential well.
	It is interesting to point out that~\cite{Liu:2021aqh}, even a truncation of the effective potential gives rise to a novel branch of QNMs that exhibits the above feature.  
	
	\section{An interplay between the spiral of the fundamental mode and echo modes}\label{sec3}
	
	In this section, we elaborate on another phenomenon related to the metric perturbation.
	As was first observed in~\cite{Cheung}, the fundamental mode spirals outward as the perturbative bump moves away from the black hole's effective potential. 
	When the effective potential is constituted by two disjoint potential barriers, such a phenomenon can be understood analytically.
	Let us denote the complex frequency of the black hole's fundamental mode by $\omega_0$.  
	It corresponds to the zero of $m_1^{++}(\omega)$, which is associated with the potential barrier $V_1(x)$.
	We assume that during the spiral, the QNM stays close to the value of $\omega_0$, namely the deviation is insignificant ($\delta\omega \approx 0$)\footnote{Note that this assumption is not valid for some instances such as the tailored P\"ochel-Teller effective potential introduced in Sec.~\ref{sec5}.}.
	Therefore, it is reasonable to approximate the Wronskian using a Taylor expansion around $\omega_0$.
	To be more specific, by maintaining the dominant terms, we have
	\begin{eqnarray}
		m_1^{++}(\omega)&\approx& m_1^{++}(\omega_0)+ m_1^{++}{'}(\omega_0) \delta \omega = m_1^{++}{'}(\omega_0) \delta \omega ,\nonumber \\
		m_2^{++}(\omega)&\approx& m_2^{++}(\omega_0) ,\nonumber \\
		m_1^{-+}(\omega)&\approx& m_1^{-+}(\omega_0) ,\nonumber \\
		m_2^{+-}(\omega)&\approx& m_2^{+-} (\omega_0) .
		\label{eq: Taylor}
	\end{eqnarray}
	In particular, we note that $m_2^{+-} (\omega_0)$ is expected to be small in magnitude, reflecting the fact that $V_2(x)$ is perturbative [c.f. the specific case given by the last line of Eq.~\eqref{eq: TransitMatrix2S}].
	Subsequently, we can reiterate Eq.~\eqref{eq: QNMs} regarding the derivation of the fundamental mode $\delta\omega$ as
	\begin{equation}
		\delta \omega \approx -e^{2 i \omega_0 L} \frac{m_1^{-+}(\omega_0)m_2^{+-}(\omega_0)}{m_1^{++}{'}(\omega_0) m_2^{++}(\omega_0)} ~.
		\label{spiralEq}
	\end{equation}
	The outward spiral can be readily explained by observing Eq.~\eqref{spiralEq}.
	As $L$ increases, the modulus and argument of the deviation $\delta \omega$ are governed by the imaginary and real parts of $\omega_0$.
	Since $\Im\omega_0 <0$, the modulus increases exponentially as a function of $L$, while the spiral rotates counterclockwise for $\Re\omega_0 > 0$.
	In particular, the spiral would not happen if $\omega_0$ is purely imaginary, which is the case for the double delta-function potential discussed earlier.
	We observe that the small magnitude of $\delta\omega$ aligns with the negligible value of $m_2^{+-} (\omega_0)$.
	For a disjoint effective potential, it is noted that similar arguments have been elaborated by a few authors~\cite{Yang:2024vor, Ianniccari:2024ysv, Qian:2024iaq}.
	Moreover, this phenomenon is not specific to the fundamental mode, as the same argument can be readily applied to other low-lying modes.
	
	Now, we argue that the spiral of the fundamental mode takes place simultaneously with the evolution of the echo modes discussed in the last section.
	Specifically, as the perturbative bump moves away from the black hole's effective potential, while the fundamental mode spirals outward from its original position, the echo modes evolve collectively, approaching the real axis.
	At a certain point, the damping rate of one of the echo modes becomes smaller than the fundamental mode and takes over the latter's place.
	
\begin{figure}[th!]
		\begin{center}
			\includegraphics[height=6cm]{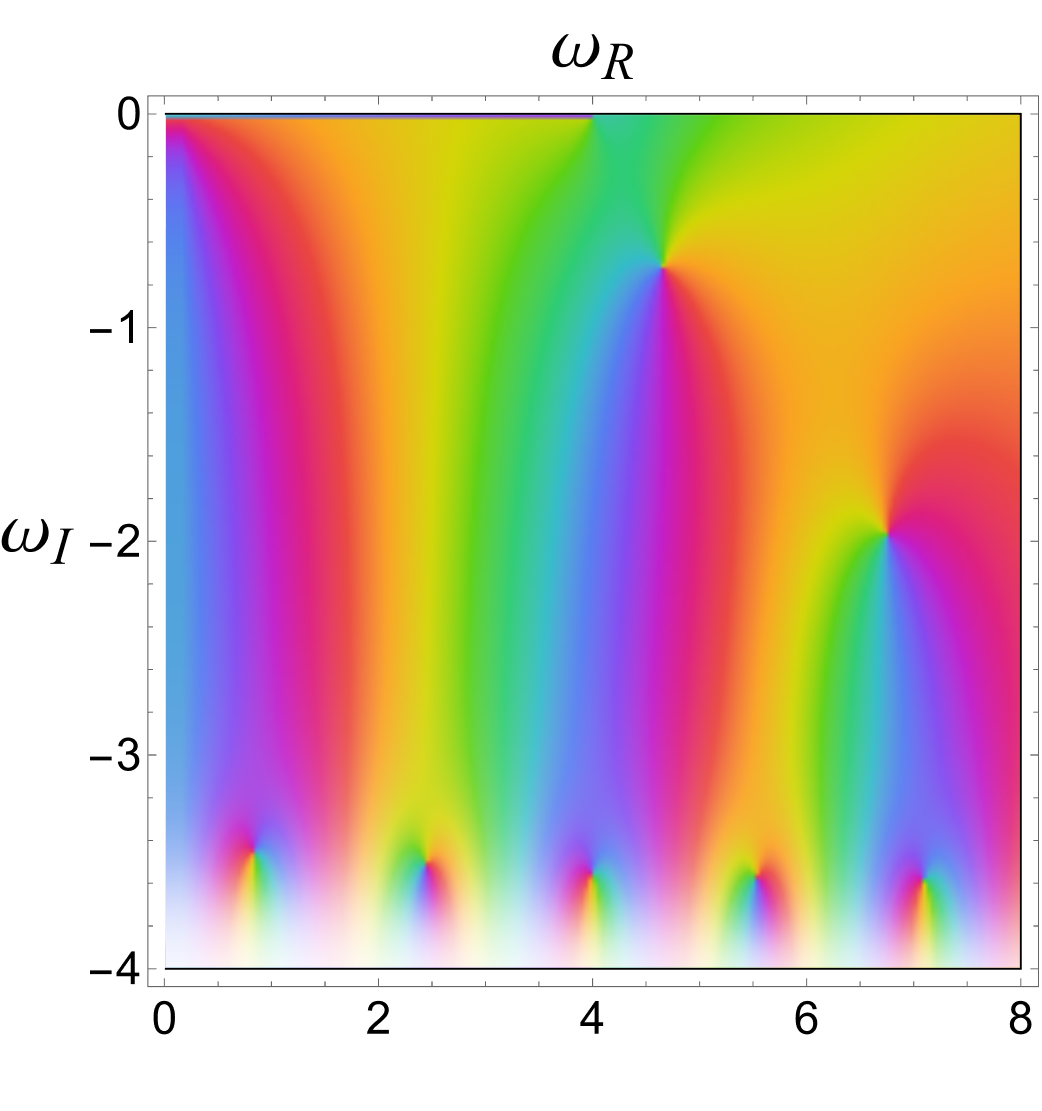}
			\includegraphics[height=6cm]{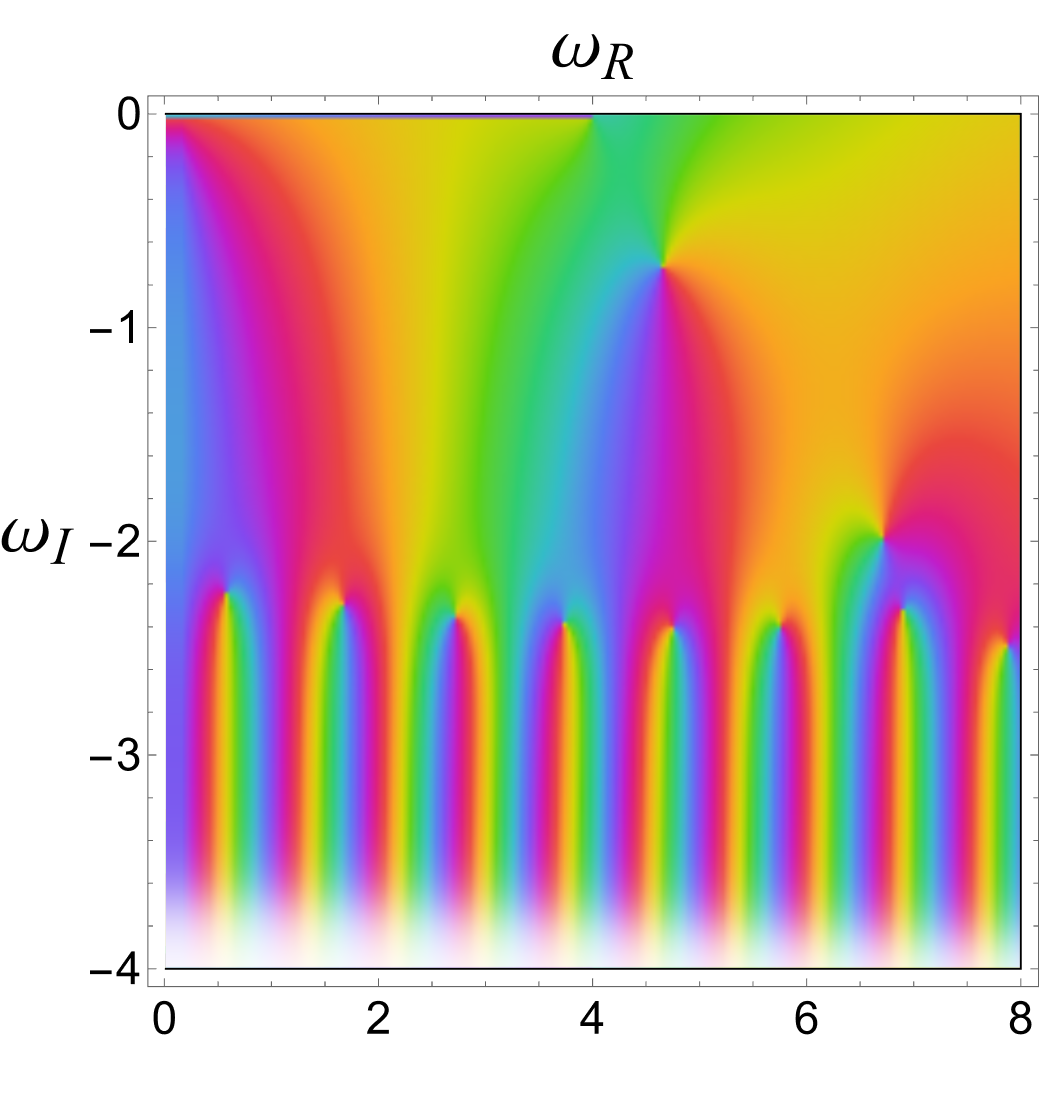}
			\includegraphics[height=6cm]{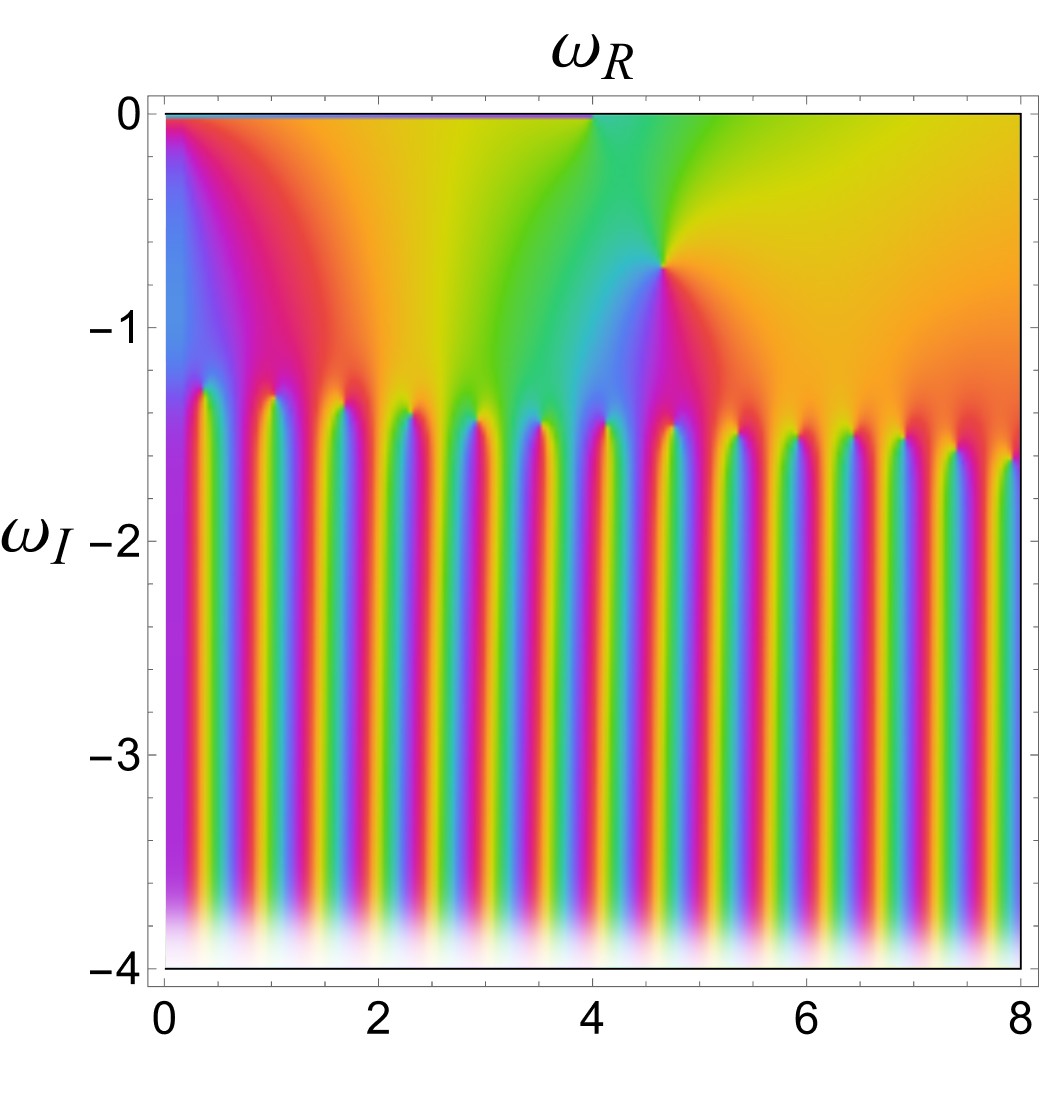}
			\includegraphics[height=6cm]{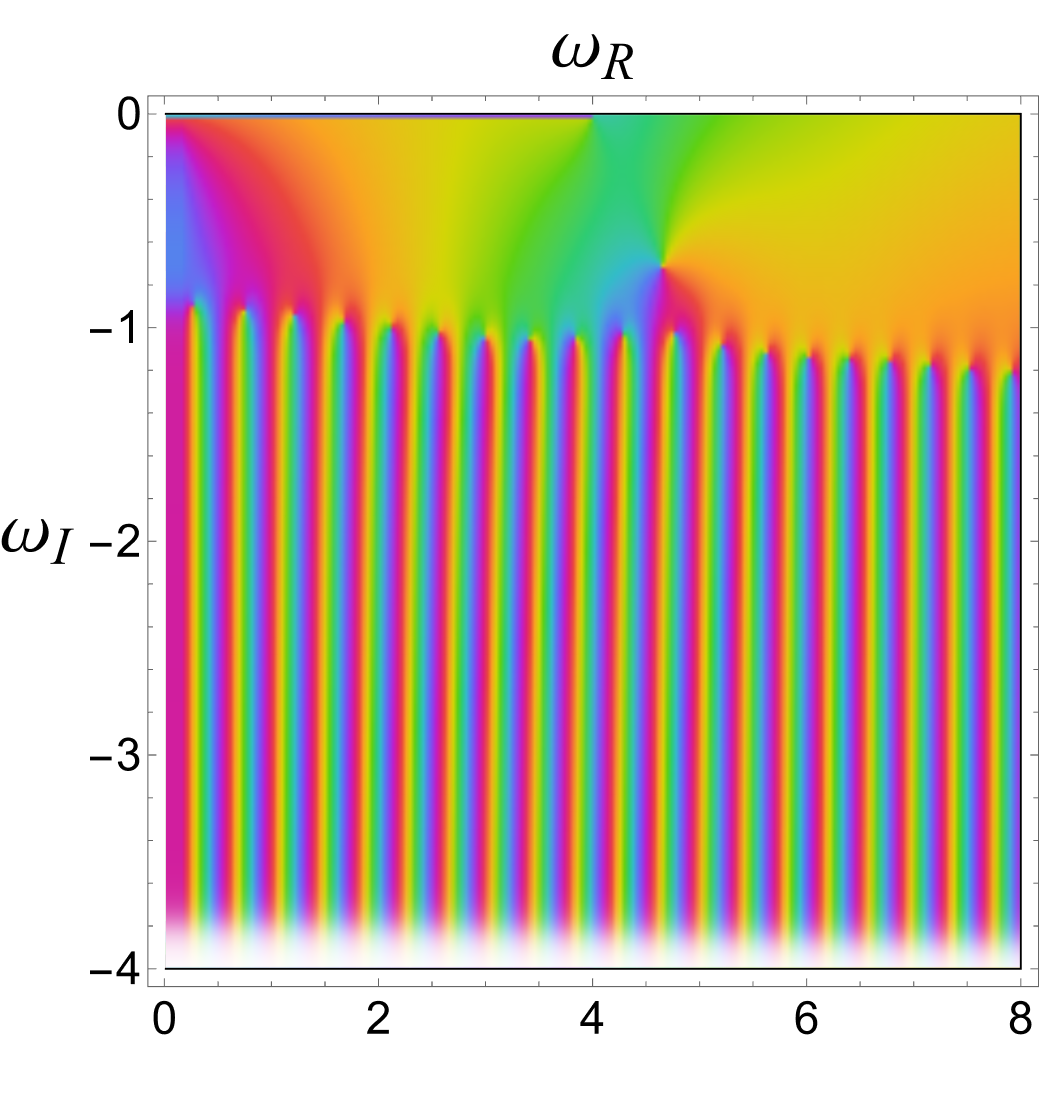}
			\includegraphics[height=6cm]{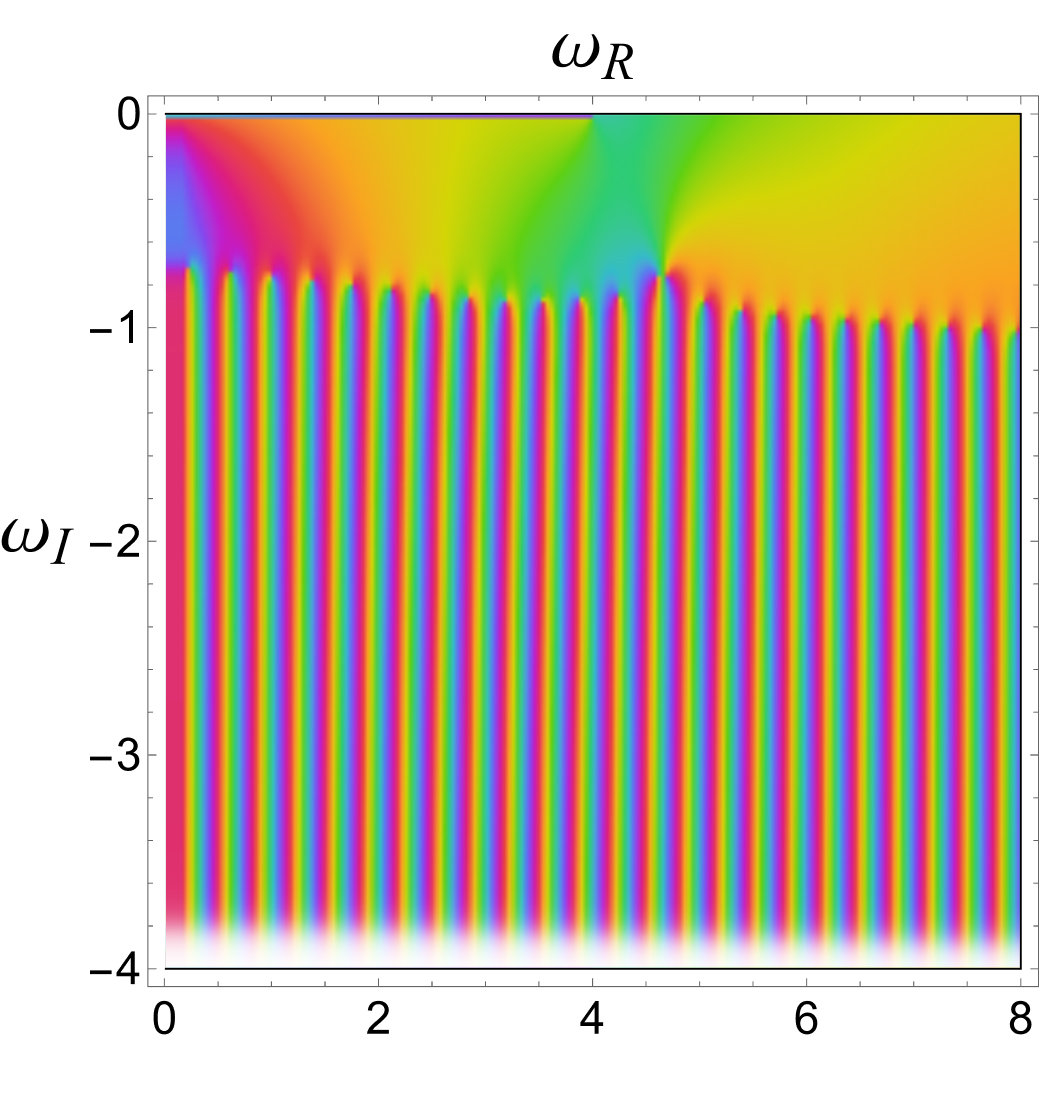}
			\includegraphics[height=6cm]{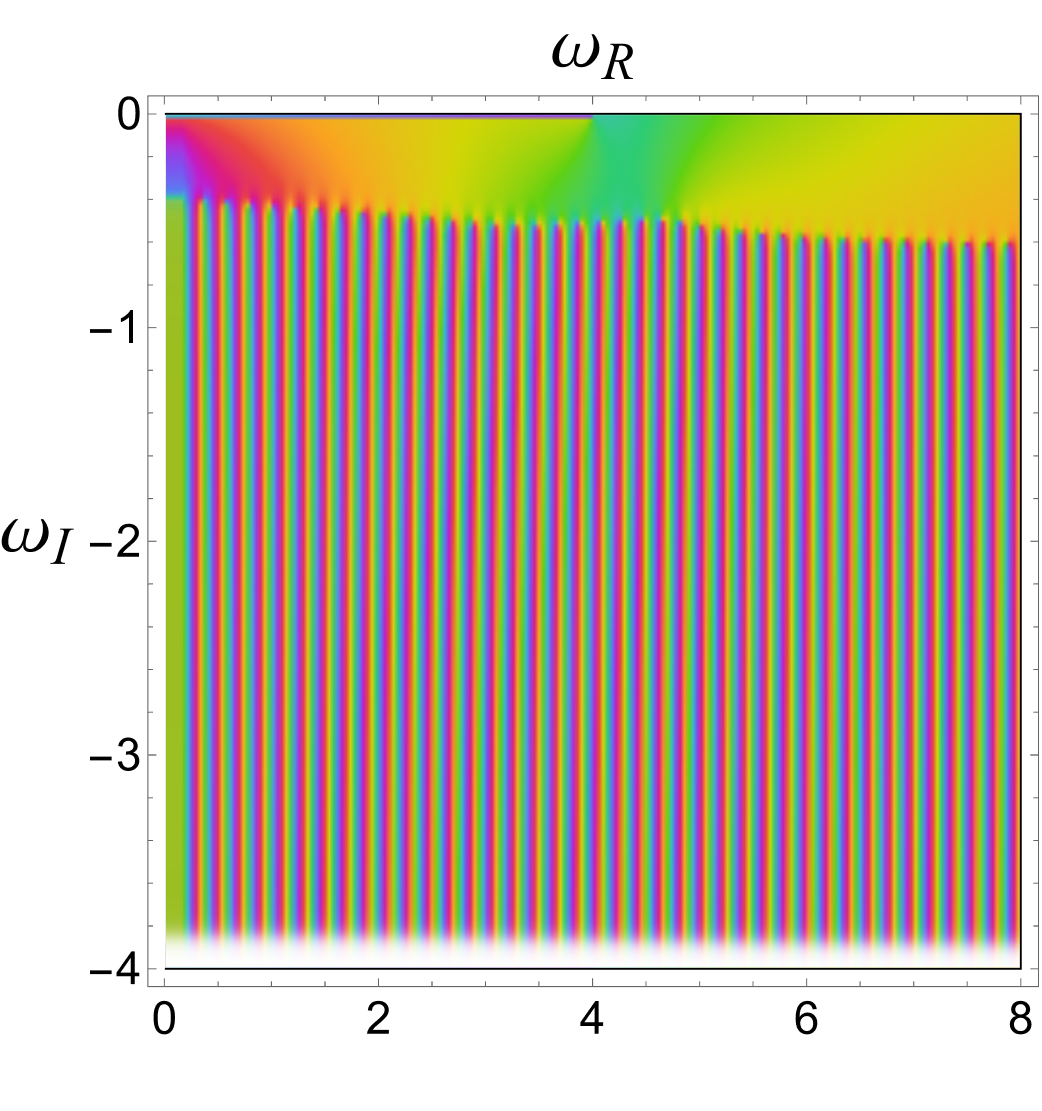}
		\end{center}
		\vspace{-0.7cm}
		\caption{\footnotesize Color map of the Wronskian Eq.~\eqref{eq: Wronskian} illustrating the evolution of the QNM spectrum in the frequency domain for toy model effective potential consisting of two potential barriers Eq.~\eqref{Vsquare}.
			The modes are represented by points encircled in red, green, and blue in a counterclockwise direction.
			In the calculations, the height of the potential is taken to be $h=16/ \sigma_0^2$, and the height and width of the bump are $\epsilon=10^{-5} /\sigma_0^2$ and $\sigma=\sigma_0$, respectively. 
			From top to bottom and left to right, we display cases for $L=2\sigma_0$, $L=3\sigma_0$, $L=5\sigma_0$, $L=7\sigma_0$, $L=8.5316\sigma_0$, and $L=15\sigma_0$.}
		\label{fig: EchoComplexPlot}
	\end{figure}

	In the literature, the taking over of the fundamental mode has been observed and discussed by several authors~\cite{Cheung, Ianniccari:2024ysv}.
	The general understanding is that the fundamental mode is taken over by one of the overtones of the original black hole's QNMs.
	Our interpretation is somewhat different from this viewpoint: the echo modes derived in the previous section do not belong to the black hole QNM spectrum but to a distinct novel branch.
	In the case of disjoint potentials, this phenomenon and its implications will be further discussed and clearly illustrated by a numerical approach presented below (see Fig.~\ref{fig: EchoComplexPlot}).
	Before delving into such discussions, we note that these findings will remain valid for more realistic scenarios where the black hole's effective potential is continuous and defined over the entire spatial domain.
	Nonetheless, as pointed out in~\cite{Cheung}, the toy model consisting of a pair of disjoint rectangular potential barriers Eq.~\eqref{Vsquare}, on which we primarily focus in the remainder of this section, reproduces many crucial features of the Regge-Wheeler potential in the presence of a perturbative bump.
	\begin{figure}[th!]
		\begin{center}
			\includegraphics[height=5cm]{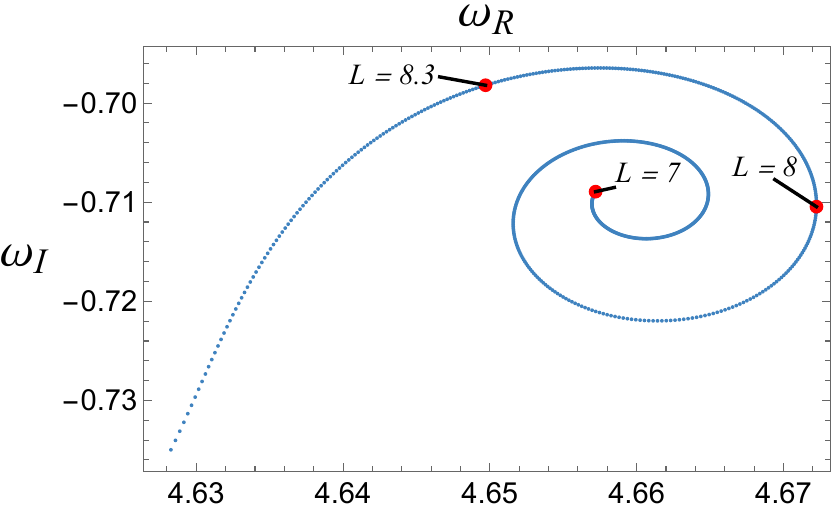}
			\includegraphics[height=5.1cm]{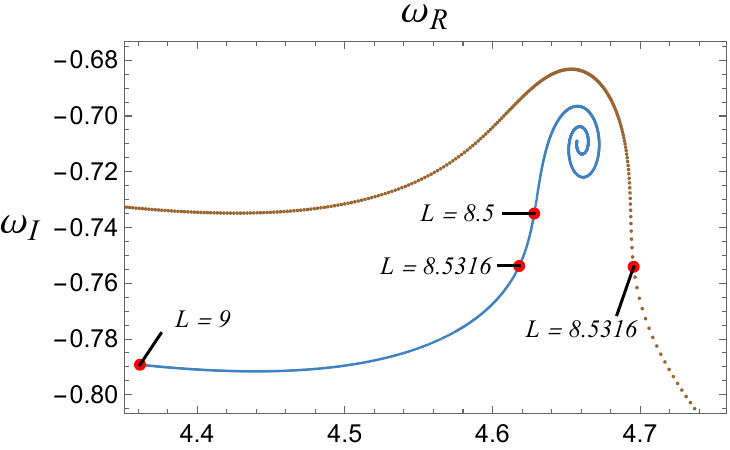}
		\end{center}
		\vspace{-0.7cm}
		\caption{\footnotesize The spiral of the fundamental QNM in the frequency domain (blue curve) is plotted for the same effective potential considered in Fig.~\ref{fig: EchoComplexPlot}.
			Different values of $L$, expressed in units of $\sigma_0$, are indicated with red dots.  
			In the right panel, we also show the motion of one of the echo modes (brown dots) that takes over the fundamental mode as $L$ increases.
			The echo mode in question moves from right to left.}
		\label{fig: Spiral}
	\end{figure}
	
	In Fig.~\ref{fig: EchoComplexPlot}, we show the evolution of the QNM spectrum in the frequency domain as the separation between the two square barriers, $L$, increases.
	Clearly, the echo modes lie primarily parallel to the real axis with similar imaginary parts.
	As $L$ increases, the modes collectively approach the real axis until they take over the fundamental mode at $L\sim 8.5316 \sigma_0$.
	Meanwhile, the distance between successive modes decreases, as was pointed out in the previous section.
	
	The spiral motion of the fundamental mode is somewhat difficult to visualize from Fig.~\ref{fig: EchoComplexPlot}.  
	This motion is explicitly illustrated in Fig.~\ref{fig: Spiral}.  
	In this figure, we continuously track the trajectory of the fundamental mode by identifying and tracing the corresponding root of the Wronskian.
	This trajectory is depicted by the blue line in the plot. 
	In the right panel of Fig.~\ref{fig: Spiral}, we also show the trajectory of the echo mode that takes over the fundamental mode in brown dots.
	Notably, the takeover occurs at $L\sim 8.5316 \sigma_0$, resulting in a jump in the real part of the frequency, indicated in the right panel of Fig.~\ref{fig: Spiral} with two red dots at $L\sim 8.5316 \sigma_0$ along the trajectories of the fundamental mode and the echo mode.  
	Comparing Fig.~\ref{fig: Spiral} with Fig.~\ref{fig: EchoComplexPlot}, the spiral motion of the fundamental mode is finite but relatively small compared to the motion of the echo modes.
	
\begin{figure}[th!]
		\begin{center}
			\includegraphics[height=5.5cm]{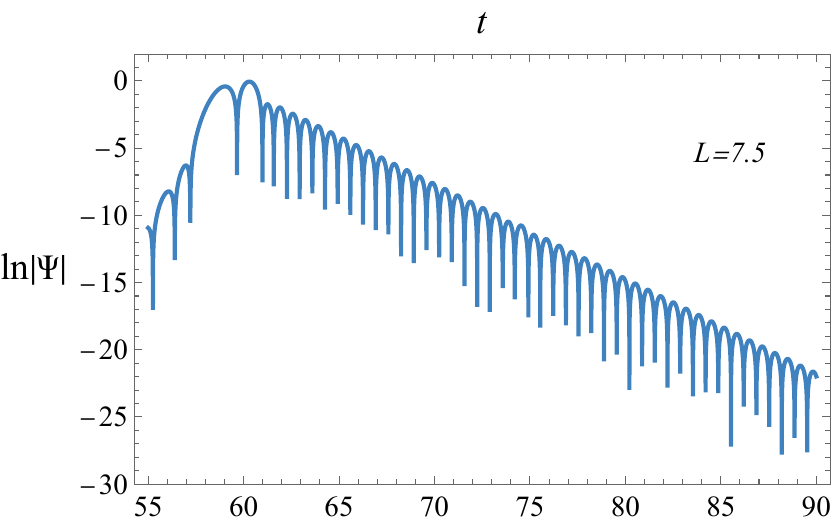}
			\includegraphics[height=5.5cm]{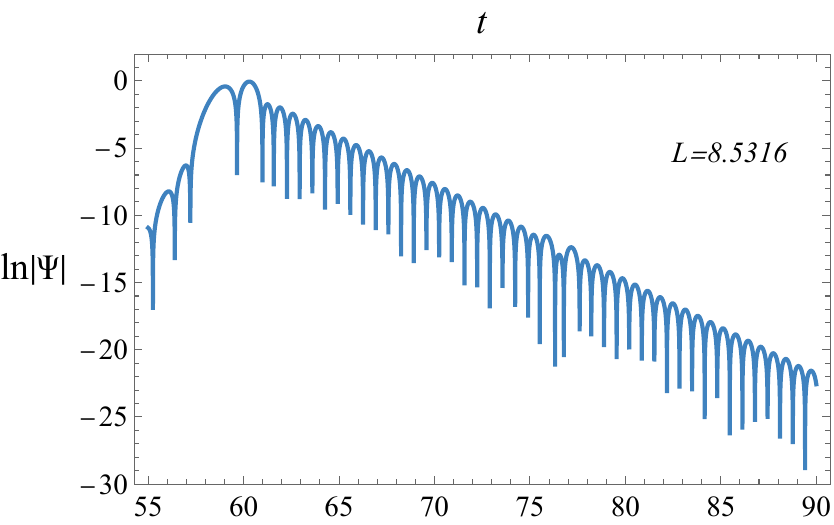}
			\includegraphics[height=5.5cm]{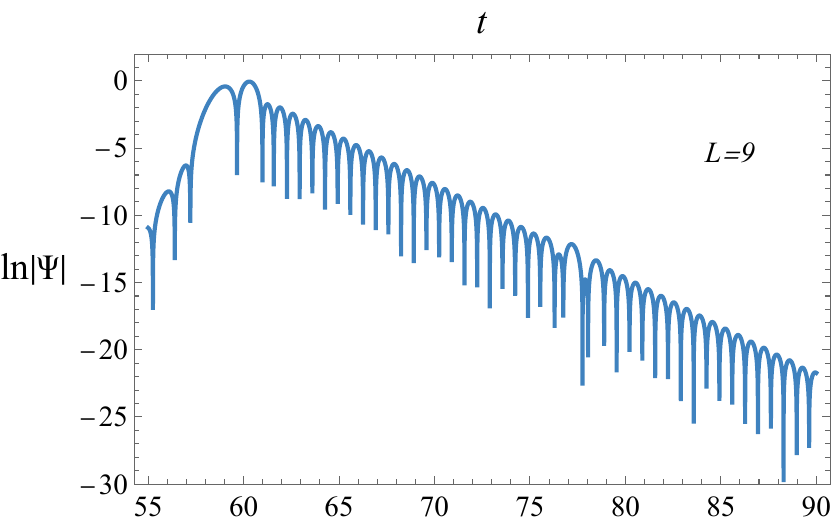}
			\includegraphics[height=5.5cm]{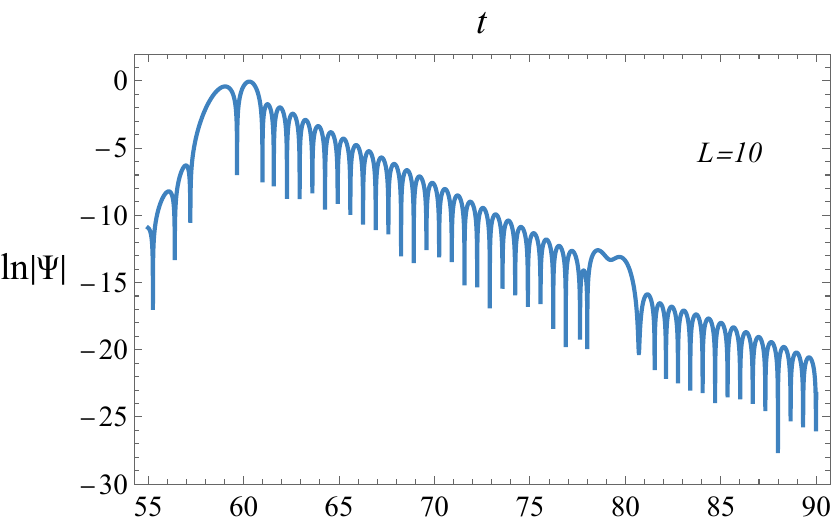}
		\end{center}
		\vspace{-0.7cm}
		\caption{\footnotesize The ringdown waveform produced by the interaction of an initial Gaussian perturbation of the form $\mathcal{A} \exp{\left[-\frac{(x-x_0)^2}{2\alpha^2}\right]}$, where $\mathcal{A}=30/\sigma_0^2$, $x_0=-10\sigma_0$, and $\alpha=\sigma_0/\sqrt{2}$,  with the same effective potential considered in Fig.~\ref{fig: EchoComplexPlot} for different values of $L$ expressed in units of $\sigma_0$.  
			The observer is located at $x=50 \sigma_0$.}
		\label{fig: EchoSeperation}
	\end{figure}

	Notably, the interplay between the fundamental mode and the echo modes in the frequency domain manifests itself in the time-domain waveforms.
	The numerical results are presented in Fig.~\ref{fig: EchoSeperation}, which displays the ringdown waveforms for various values of $L$.
	We observe that the takeover of the fundamental mode, marking the end of its spiral motion, coincides with the moment when the echo waveform becomes distinguishable from the underlying ringdown waveform.
	Specifically, Fig.~\ref{fig: EchoSeperation} shows that the first echo begins to merge around $L\sim 8.5316 \sigma_0$, exactly where the echo mode overtakes the fundamental QNM.
	This phenomenon can be readily explained:
	On the one hand, the time-domain quasinormal oscillations are primarily dominated by the fundamental QNM, with the ringdown's damping rate governed by the imaginary part of the fundamental frequency.
	On the other hand, when a series of modes lie parallel to the real axis, the waveform essentially exhibits echoes~\cite{Liu:2021aqh, Qian:2021aju}.
	The takeover moment signifies that echo modes start to play a more important role in the time-domain waveform than the fundamental mode.
	The echoes appear slightly before the takeover occurs because these modes act collectively, and therefore, their effect becomes more pronounced earlier than expected.
	Consequently, the time-domain waveform transitions from a purely damped wave, dominated by the fundamental mode's damping rate, to a waveform enveloped by echoes right about the takeover process.
	
	\section{Generalization to the continuous case}\label{sec4}
	
	To generalize the results developed in Secs.~\ref{sec2} and~\ref{sec3} to the case of a continuous effective potential, we start by demonstrating the spirit of our arguments via explicit calculations.
	Specifically, we mimic the black hole metric by a P\"oschl-Teller potential where the perturbation is introduced by a square barrier.
	More specifically, we consider the following effective potential
	\begin{equation}
		V(x) = \left\{ \begin{array}{lll}
			V_\mathrm{PT}(x)  &~~~~~& x \le L-\frac{\sigma}{2}~\\    \\
			V_\mathrm{PT}\left(L-\frac{\sigma}{2}\right)+\epsilon &~~~~~& L-\frac{\sigma}{2} < x < L+\frac{\sigma}{2}~\\    \\
			V_\mathrm{PT}(x) &~~~~~& x \ge L+\frac{\sigma}{2}~
		\end{array}
		\right. ,    
		\label{VPTsquare2}
	\end{equation}
	where 
	\bqn
	V_\mathrm{PT}(x)=\frac{V_0}{\cosh^2(\kappa x)} \lb{PTpotential}
	\eqn
	with $\epsilon \ll V_0$ and $L\gg 1$.
	In Eq.~\eqref{VPTsquare2}, the metric perturbation is implemented by adding a minor square barrier on top of a P\"oschl-Teller effective potential far from the black hole.
	
	To evaluate the Wronskian, it is not difficult to show that one can safely ignore the ingoing part of the asymptotic outgoing wave in the region $x \ge L+\frac{\sigma}{2}$, due to the properties of the hypergeometric functions~\cite{Qian:2024iaq}.
	Subsequently, the Wronskian for the potential reads
	\begin{equation}
		\begin{array}{ll}
			e^{i\sigma (\omega -\tilde{\omega})} (1+e^{2\kappa L-\kappa \sigma})^\beta\Bigl\{ 2 i \beta \kappa \left[\omega+\tilde{\omega}-e^{2 i \sigma \tilde{\omega}}(\omega-\tilde{\omega})\right]
			~   _2F_1(1+\beta, \beta-i\omega/\kappa, 1-i\omega/\kappa; -e^{2\kappa L-\kappa \sigma}) \Bigr. \\
			+\frac{\left[\omega (1-e^{2i \sigma \tilde{\omega}})\left(\omega e^{2\kappa L}+e^{\kappa \sigma}(\omega-2i \beta \kappa)\right) +2 \tilde{\omega} (1+e^{2i \sigma \tilde{\omega}})\left(\omega e^{2\kappa L}+e^{\kappa \sigma}(\omega-i \beta \kappa)\right) +\tilde{\omega}^2 (e^{2\kappa L}+e^{\kappa \sigma})(1-e^{2i \sigma \tilde{\omega}})\right]}{e^{2\kappa L}+e^{\kappa \sigma} }\\
			\Bigl. ~~~\times ~ _2F_1(\beta, \beta-i\omega/\kappa, 1-i\omega/\kappa; -e^{2\kappa L-\kappa \sigma})\Bigr\},
			\label{Wrons2}
		\end{array}  
	\end{equation}
	where $\tilde{\omega}=\sqrt{\omega^2-\left[V_\mathrm{PT}\left(L-\frac{\sigma}{2}\right)+\epsilon\right]}$ and $\beta=(1+\sqrt{1-4V_0/\kappa^2})/2$. 
	The condition for a vanishing Wronskian can be simplified by taking into account $L\gg 1$, the asymptotic properties of hypergeometric functions, and $\omega_R\gg |\omega_I| \gg 1$, which gives
	\bqn
	\mathbf{f}(\omega) \approx  \frac{2 i \omega^2 e^{\pi \omega / \kappa}}{ \tilde{\epsilon} \sin(\pi \beta)e^{2 i \sigma \omega}}.
	\lb{asymptoticRegionFSimplified}
	\eqn
	The details of the above derivations are relegated to Appx.~\ref{app1}.
	
	The modulus of $\mathbf{f}(\omega)$ has the form 
	\begin{equation}
		f(\omega)\equiv |\mathbf{f}(\omega)|\approx  \frac{2  \omega_R^2 e^{\pi \omega_R / \kappa}}{ \tilde{\epsilon} |\sin(\pi \beta)|e^{2  \sigma |\omega_I|}}.
	\end{equation}
	Subsequently, at the high overtone limit $n\gg 1$, to the lowest order, we have
	\bqn
	\omega_R &\approx & \left(n+\frac12\right)\frac{\pi}{L},\nb\\
	\omega_I &\approx& -\frac{1}{2L} \left[ \frac{\pi \omega_R}{\kappa} +2\ln{\omega_R} \right].
	\label{eq: PT-QNM-asymp}
	\eqn
	The above results are very similar to what we obtained for the disjoint potentials.
	
\begin{figure}[th!]
		\begin{center}
			\includegraphics[height=6.cm]{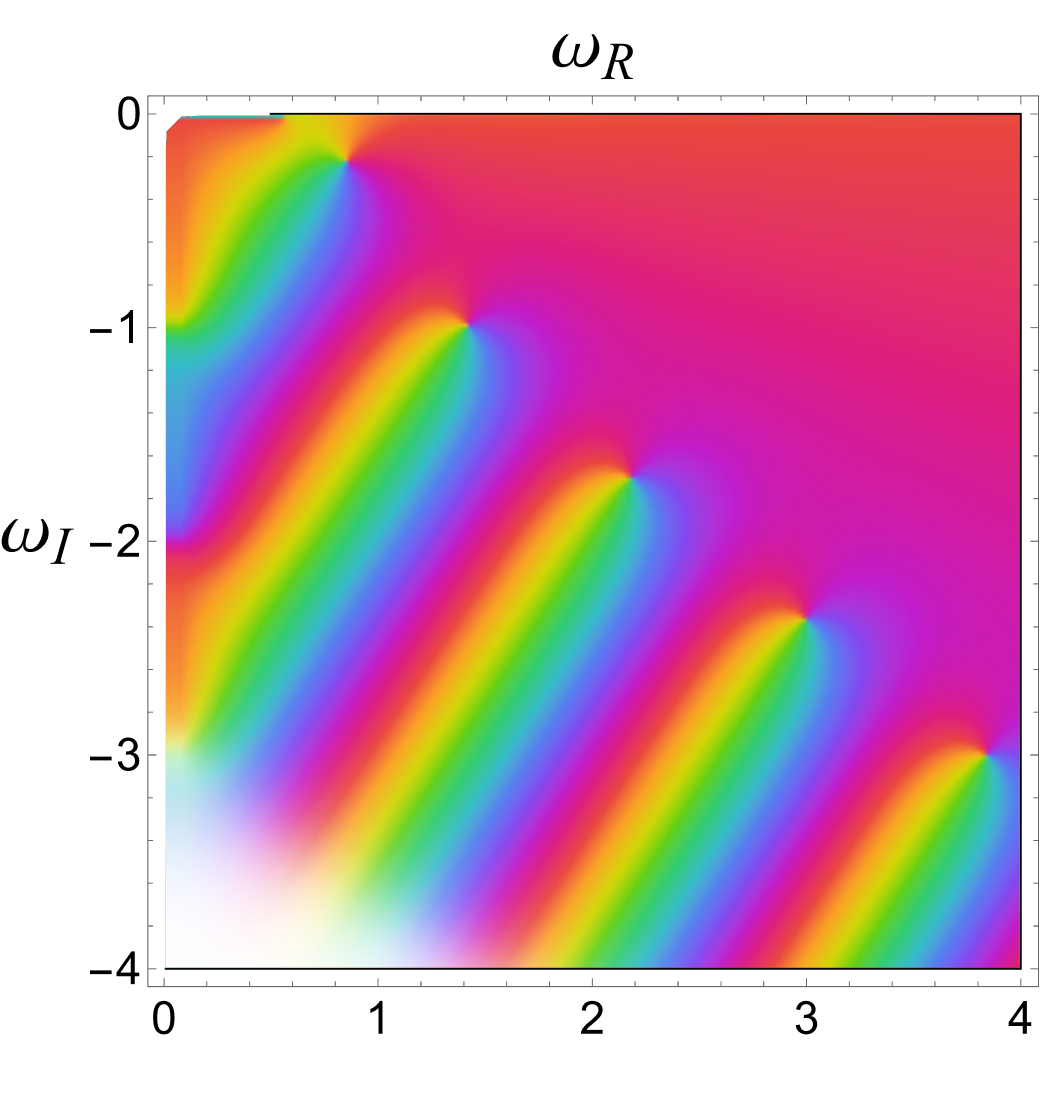}
			\includegraphics[height=6.cm]{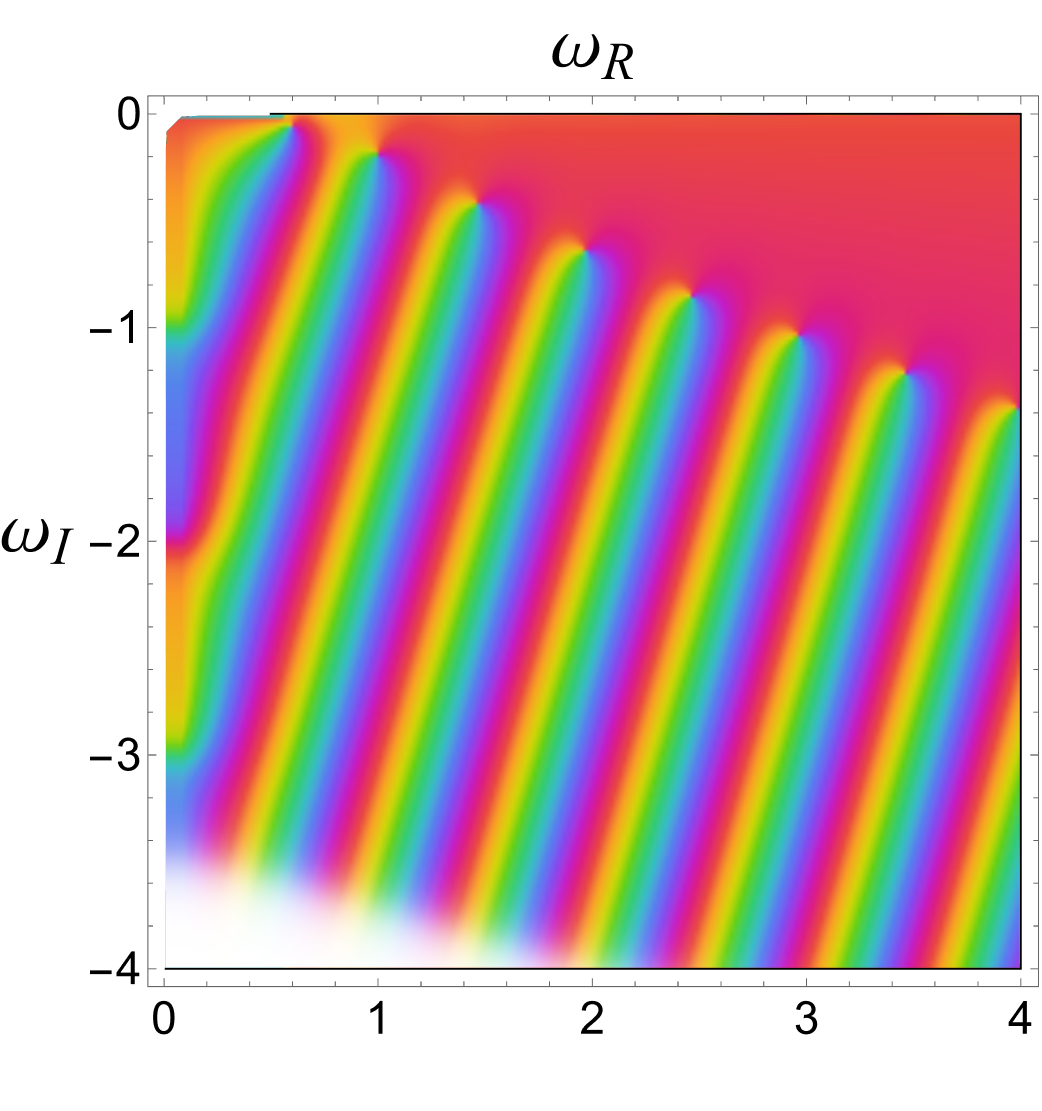}
			\includegraphics[height=6.cm]{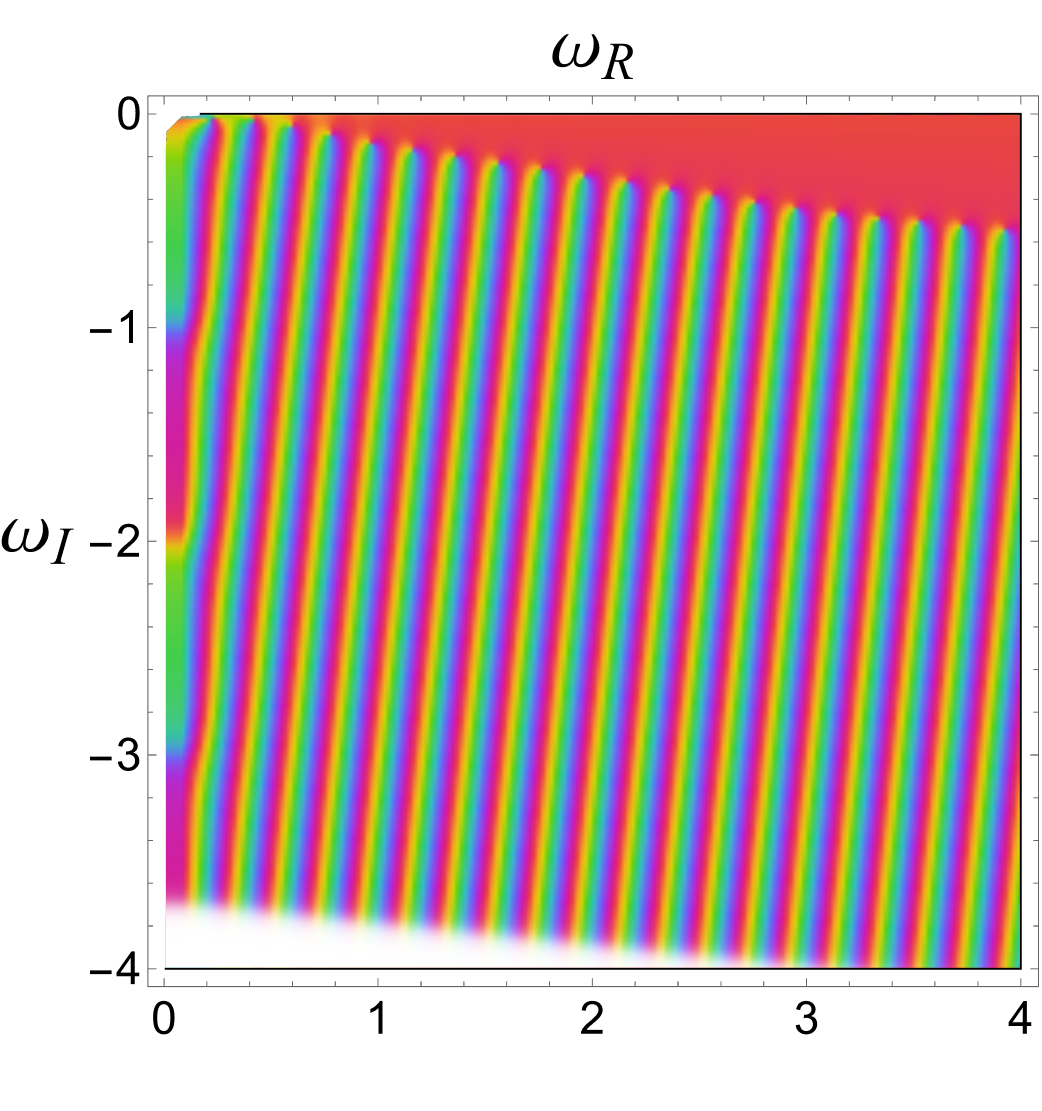}
		\end{center}
		\vspace{-0.7cm}
		\caption{\footnotesize The QNM spectrum for the effective potential Eq.\ \eqref{VPTsquare2}, where $V_0=1/\sigma^2$, $\kappa=1/\sigma$, and $\epsilon=0.05/\sigma^2$. The modes are represented by points encircled in red, green, and blue in a counterclockwise direction.
			From left to right, $L=2\sigma$, $L=5\sigma$, and $L=15\sigma$, respectively.}
		\label{fig: fundamentalModeSpin}
	\end{figure}

	In Fig.~\ref{fig: fundamentalModeSpin}, we show how the resulting QNMs of the potential Eq.~\eqref{VPTsquare2} move in the complex plane as $L$ increases.
	There are qualitative differences between this case and the disjoint potential shown in Fig.~\ref{fig: EchoComplexPlot}. 
	For the latter, the echo modes constitute a novel branch of QNMs.
	As discussed in the previous section, the echo modes gradually take over the original black hole's QNM spectrum until the fundamental mode is eventually taken over. 
	In the case of potential Eq.~\eqref{VPTsquare2}, there is only one branch.  
	As $L$ increases, the QNM spectrum deforms and gradually lies parallel to the real axis, as elaborated in~\cite{Li:2024npg}.
	Although Eq.~\eqref{eq: PT-QNM-asymp} does not provide such a ``dynamic'' picture, it is consistent with the numerical results in Fig.~\ref{fig: fundamentalModeSpin}.
	
	\section{Causality dilemma and feasibility of black hole spectroscopy involving echo modes}\label{sec5}
	
	As discussed in the {\it Introduction}, the instability of the QNM spectrum introduces a causality-related dilemma.  
	The authors of~\cite{Hui} have attempted to resolve this causality issue using the frequency domain.  
	It is shown in the study that the Green's function for two disjoint bounded potentials has the general form
	\bqn
	G(t,x|t',x') &=& \int_{c-i\infty}^{c+i\infty}\frac{ds}{2\pi i}e^{s(t-t')}\frac{1}{2s(m_1^{++} m_2^{++}+ e^{-2 s L} m_2^{+-}m_1^{-+})}\nb\\
	&&(\phi_-(s,x)\phi_+(s,x')\theta(x-x')+\phi_+(s,x)\phi_-(s,x')\theta(x'-x))~,
	\label{eq: GreenFunc}
	\eqn
	where $s=-i\omega$, $\phi_\pm$ are two linearly-independent solutions to the wave equation, and $c$ is a small positive real number that serves the purpose of making sure the integration contour includes all the roots of the Wronskian in the complex plane.  
	The authors then Taylor expand the above expression to obtain
	\bqn
	G(t,x|t',x') &=& \sum_{k=0}^{\infty}(-1)^k\int_{c-i\infty}^{c+i\infty}\frac{ds}{2\pi i}\frac{(m_2^{+-}m_1^{-+})^k}{2s(m_1^{++} m_2^{++})^{k+1}} e^{s(t-t'-2kL)} \nb\\
	&&(\phi_-(s,x)\phi_+(s,x')\theta(x-x')+\phi_+(s,x)\phi_-(s,x')\theta(x'-x)).
	\label{eq: GreenFunc}
	\eqn
	Note that the sum truncates when $2kL > t-t'$.  
	The authors conclude that ``the sum over $k$ can be understood as a sum over the number of back-and-forth bounces between the two potentials. 
	Causality is encoded in the sum, and at any finite time, the sum truncates to the maximum number of bounces permitted by causality.
	Over sufficiently short times, causality forbids a back-and-forth bounce, and one would only observe the quasinormal frequencies associated with the individual potentials.''   
	
	In this regard, if one strictly follows Hui \textit{et al.}'s expansion of the Laplace transformed Green's function Eq.~\eqref{eq: GreenFunc}, the echo modes do not correspond to any pole of the Green's function.  
	This claim can be elaborated as follows.  
	Eq.~\eqref{eq: GreenFunc} involves a geometric series of the form
	\bqn
	S=a_1+a_2+\cdots = a_1 +a_1q+a_1 q^2\cdots ,\lb{InfSum}
	\eqn
	where
	\bqn
	a_1 &=& \frac{1}{4\pi i s}\frac{1} {m_1^{++}m_2^{++}}e^{s(t-t')},\nb\\
	q &=& -\frac{m_2^{+-}m_1^{-+}}{m_1^{++}m_2^{++}}e^{-2sL}\propto e^{-2sL}=e^{2i\omega L} .\lb{Def_a1q}
	\eqn
	For a finite sum, we have
	\bqn
	S_m = a_1+a_1 q+a_1 q^2+\cdots+a_1 q^{m-1}= a_1\frac{1-q^m}{1-q}.
	\eqn
	Despite the appearance of the term $(1-q)$ in the denominator, the above expression does not have a pole at $q\sim 1$.
	This is because the numerator contains the factor $(1-q^m)$. 
	In other words, as long as only a finite number of terms are involved, the pole structure of the Green's function is precisely the same as that of the original black hole, governed by $\left[ m_1^{++}(\omega) m_2^{++}(\omega)\right]$ in the denominator of $a_1$.  
	
	Following this line of argument, a simple yet illustrative counterexample is the effective potential constituted by the two disjoint delta functions Eq.~\eqref{twoDelta}.
	In this case, since each potential only hosts a single QNM, the denominator of $a_1$ defined in Eq.~\eqref{Def_a1q} implies two poles at $\omega_{1,2}=-\frac12 iv_{1,2}$.  
	This immediately leads to a dilemma, as it is evident that the ringdown waveform and the subsequent echoes cannot be constructed from merely two modes.
	Moreover, analyses associated with spectral instability indicate the evolution or emergence of novel QNMs.
	Specifically, as explored in~\cite{Li:2024npg, Qian:2024iaq}, a deformation of the QNM spectrum is identified.  
	
	To resolve this dilemma, one must include an infinite number of terms by considering the infinite sum Eq.~\eqref{InfSum} that has the compact form
	\bqn
	S=\frac{a_1}{1-q},
	\eqn
	which allows one to recover the roots of the Wronskian Eq.~\eqref{eq: Wronskian} that contains all the echo modes.  
	Therefore, we emphasize that the echoes are a collective phenomenon, demonstrated by the fact that the poles associated with the echoes do not emerge unless we include an infinite number of terms in the sum.
	
\begin{figure}[th!]
		\begin{center}
			\includegraphics[height=5.5cm]{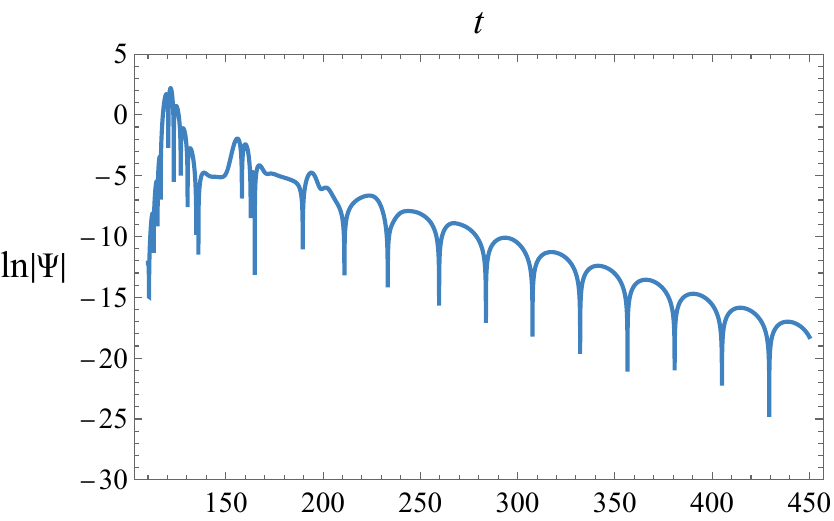}
			\includegraphics[height=5.5cm]{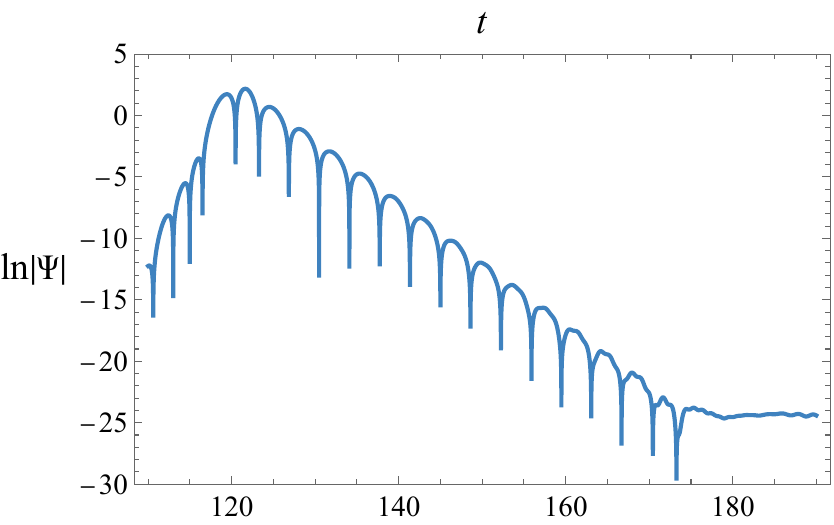}
		\end{center}
		\vspace{-0.7cm}
		\caption{\footnotesize 
			The ringdown waveform produced by an initial Gaussian perturbation of the form $\mathcal{A} \exp{\left[-\frac{(x-x_0)^2}{2\alpha^2}\right]}$, where $\mathcal{A}=18/\sigma^2$, $x_0=-20\sigma$, and $\alpha=\sigma$ for different effective potentials. The observer is located at $x=100 \sigma$.
			Left: The modified P\"oschl-Teller potential Eq.~\eqref{VPTsquare2} with $V_0=1/\sigma^2$, $\kappa=1/\sigma$, $L=18\sigma$, and $\epsilon=0.05\sigma$. 
			Right: The original P\"oschl-Teller potential Eq.~\eqref{PTpotential} with the same parameters.}
		\label{fig: EchoWaveform}
	\end{figure}

\begin{figure}[th!]
		\begin{center}
			\includegraphics[height=5.5cm]{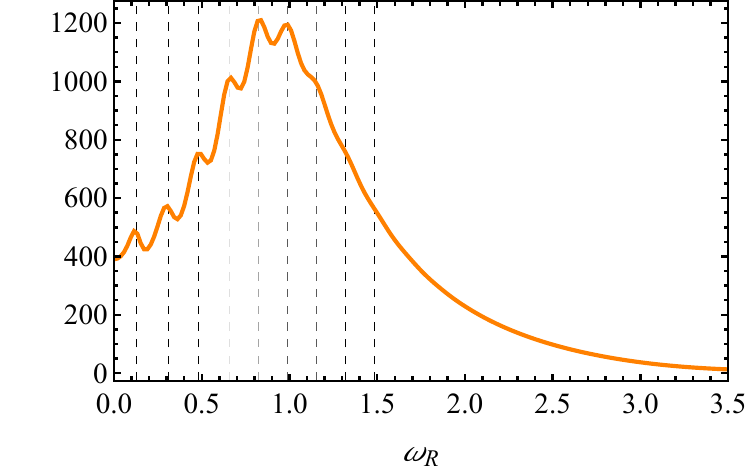}
			\includegraphics[height=5.5cm]{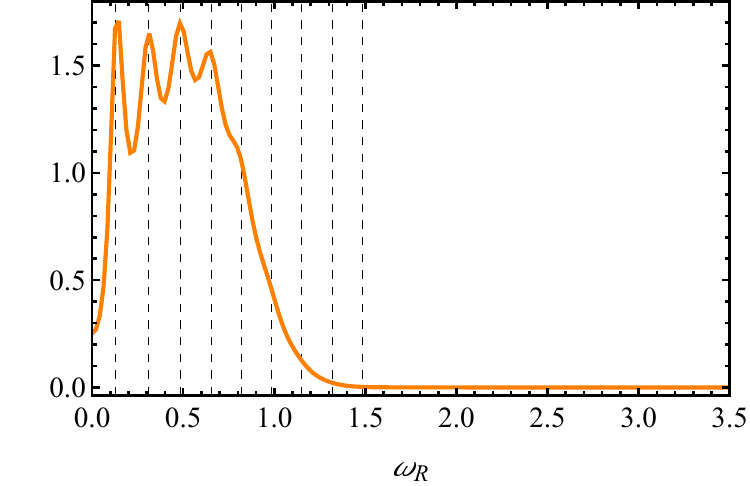}
			\includegraphics[height=5.5cm]{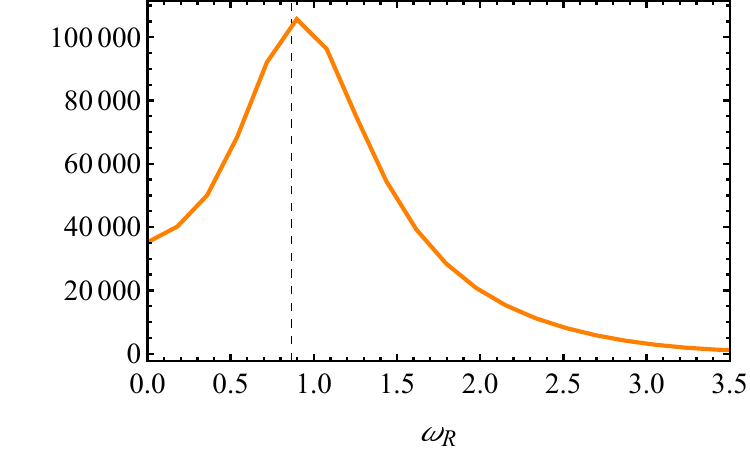}
		\end{center}
		\vspace{-0.7cm}
		\caption{\footnotesize The Fourier transformed frequency profiles of the ringdown waveform shown in the left panel of Fig.~\ref{fig: EchoWaveform}.
			The analysis is carried out for different time intervals, where time is expressed in units of $\sigma$.
			The gray lines indicate the location of the real part of the echo modes.
			Top left: One considers the interval $120\sigma\le t \le 450\sigma $, which includes the entire time domain.
			Top right: One considers the interval $ 150\sigma\le t \le 450\sigma $, where the initial ringdown has been excluded.
			Bottom: One considers the interval $120\sigma\le t \le 155\sigma $, where the echo pulses are excluded.  
			In the bottom panel, the jagged profile in the frequency domain is a direct consequence of the limited time interval used for analysis, which leads to fewer discrete frequency values in the Fourier transform.
			For a longer time interval, the Fourier transform curve becomes smoother, as shown in Fig.~\ref{fig: Fourier}.}
		\label{fig: EchoFourier}
	\end{figure}

\begin{figure}[th!]
		\begin{center}
			\includegraphics[height=5.5cm]{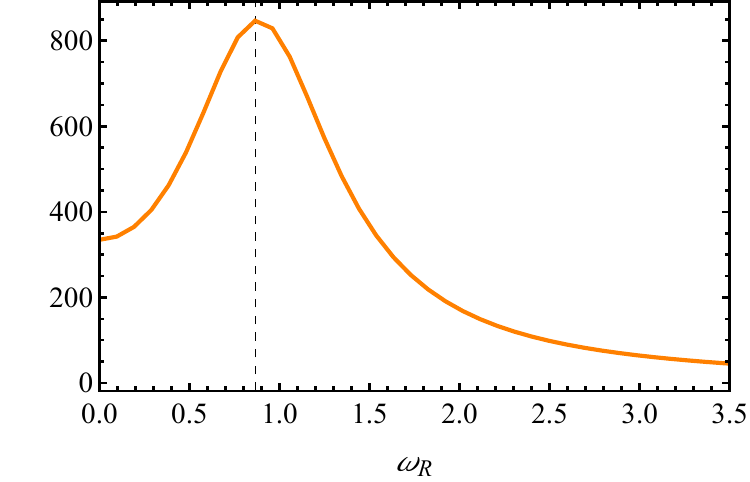}
		\end{center}
		\vspace{-0.7cm}
		\caption{\footnotesize The Fourier transformed frequency profile of the ringdown waveform shown in the right panel of Fig.~\ref{fig: EchoWaveform}.
			The gray line indicates the location of the real part of the fundamental QNM of the P\"oschl-Teller potential Eq.\ \eqref{PTpotential} at $\omega_R = \sqrt{V_0-\kappa^2/4}=0.866/\sigma$.}
		\label{fig: Fourier}
	\end{figure}

	One might notice that the above arguments, along with the calculations presented in~\cite{QNM-Significance1, QNM-Significance2, Cheung, Li:2024npg, Qian:2024iaq}, reside primarily in the frequency domain.
	To confirm that such echo poles are not merely a theoretical artifact, it is meaningful to extract them by analyzing the time-domain waveforms.
	Indeed, a Fourier waveform analysis shows that one can identify the echo modes \cite{Bueno}.
	In the left panel of Fig.~\ref{fig: EchoWaveform}, we show the ringdown waveform produced by the interaction of a Gaussian wavepacket with the potential given in Eq.~\eqref{VPTsquare2}.
	For comparison, in the right panel of Fig.~\ref{fig: EchoWaveform}, we show the corresponding waveform of the original P\"oschl-Teller potential Eq.~\eqref{PTpotential}.
	In Figs.~\ref{fig: EchoFourier} and~\ref{fig: Fourier}, we show the corresponding Fourier transform of the ringdown waveforms.
	The left panel of Fig.~\ref{fig: EchoFourier} gives the frequency domain profile obtained from the waveform shown in the left panel of Fig.~\ref{fig: EchoWaveform}.
	Fourier analysis in Fig.~\ref{fig: EchoFourier} is carried out for three different time intervals by considering the entire range (top left panel), excluding the initial ringdown (top right panel), and finally excluding the echoes (bottom panel).
	It is noted that frequency takes discrete values because the time-domain waveform is a time series evaluated over a finite interval.
	Specifically, the length of the time interval governs the resolution of the frequency-domain profile, while the time series sampling rate determines the relevant range of the frequency axis.
	In the top left and top right panels, the real parts of the echo modes mostly match the peaks in the wiggled profile, indicating that one can successfully extract the echo modes.\footnote{In practice, the Fourier transform slightly varies depending on the specific choice of the time interval.  
		We argue that this issue does not change the qualitative conclusions of our paper.
		However, to employ the Fourier transform as a potential observational tool, further investigation is necessary to resolve such uncertainty entirely.} 
	Furthermore, by comparing the top left and top right panels, it is interesting to point out that the overall shape of the profile shown in the top left panel is similar to the Fourier transform curve in Fig.~\ref{fig: Fourier} generated from the ringdown waveform of the original P\"oschl-Teller potential shown in the right panel of Fig.~\ref{fig: EchoWaveform}.
	This indicates that while the fundamental mode is taken over and does not appear unambiguously in the QNM spectrum, it persists by enveloping the peaks of the echo modes.
	Conversely, although the echo modes are again observed in the top-right panel of Fig.~\ref{fig: EchoFourier}, the shape of the curve no longer resembles the profile in Fig.~\ref{fig: Fourier}. 
	This is understood as the black hole's fundamental frequency having a more significant magnitude of the imaginary part than those of the echo modes, and subsequently, it leaves less impact on the resulting Fourier transform.
	Lastly, in the bottom panel of Fig.~\ref{fig: EchoFourier}, as echoes are excluded from the analysis, one recuperates the original black hole's fundamental mode, in agreement with the profile shown in Fig.~\ref{fig: Fourier}.
	It is noted that the bottom panel of Fig.~\ref{fig: EchoFourier} features a jagged profile.
	This is understood as a result of the limited time interval used for the analysis, leading to fewer discrete frequency values in the Fourier transform. 
	In comparison, using a longer time interval, the Fourier transform curve becomes smoother, as shown in Fig.~\ref{fig: Fourier}.
	
	Finally, we would like to elaborate more on the causality dilemma by pointing out that the results obtained in this section show that the Fourier transform of the ringdown changes shape as one gradually receives more signals from the source. 
	If an observer receives and Fourier transforms the initial ringdown shown in the left panel of Fig.~\ref{fig: EchoWaveform},  they will find the profile in the bottom panel of Fig.~\ref{fig: EchoFourier} and determine a frequency consistent with the fundamental QNM of an unperturbed effective potential.
	As the observer receives more signals (i.e., waveforms that include at least the first echo pulse), the Fourier transform begins to resemble the profile shown on the left panel of Fig.~\ref{fig: EchoFourier}.
	This seemingly indicates that the QNM spectrum evolves as the observer gathers more information about the signal.
	In theory, QNMs are well defined in the frequency domain and do not evolve.
	This apparent evolution is simply due to a numerical implementation of the Fourier transform on a finite time interval. In principle, a Fourier transform must encompass the entire time domain rather than a finite time interval.
	The above analysis, conducted for a localized/disjoint perturbative bump, presents the causality dilemma in its most pronounced form, allowing for semianalytical treatment.
	For realistic scenarios with continuous potentials, distinguishing echo pulses from quasinormal oscillations might not be feasible, and conclusions drawn from such toy models may not generalize straightforwardly.
	In practice, however, this indeed presents a genuine challenge for data analysis.
	Nevertheless, as noted by many authors, the time-domain waveform remains largely intact~\cite{Nollert1, DGM-significance, Qian-QNM-Significance, Ringdown-Stability}, even in the presence of significant spectral instability.
	However, this ``time-domain stability'' should not be readily interpreted as stability in the context of black hole spectroscopy.
	This caution is warranted because our goal is to extract the underlying metric parameters, and similar time-domain waveforms might be ``deceptive'', potentially concealing essential information.

	\section{Concluding remarks}\label{sec7}
	
	This study investigated echo modes and their interplay with the black hole's original QNMs.
	By explicit analytic calculations, we explored the properties of the echo modes whose emergence is due to metric perturbations manifesting as minor discontinuity/bump in both disjoint and continuous black hole effective potentials.
	These echo modes feature a distribution primarily parallel to the real frequency axis.
	As the bump moves away from the black hole horizon, the echo modes collectively move closer to the real axis while the distance between successive modes decreases.  
	We showed that this behavior is universal.
	In the frequency domain, such an evolution of the echo modes may interact with the out-spiral of the black hole's fundamental QNM.
	Subsequently, this leads to a jump in the real part of the fundamental mode at the moment when the original fundamental mode is taken over by one of the echo modes with a marginally lower damping rate.
	In the time domain, the resulting waveform transits from primarily damped oscillations, dominated by the fundamental mode, to echo waves, characterized by periodic echo pulses that can be essentially constructed by echo modes with lower damping rates.  
	It is argued that such an interaction between the evolution of echo modes and possible out-spiral of black hole QNMs is universal, further complementing the recent findings~\cite{Qian:2024iaq}.
	
	Furthermore, we elaborated on the causality dilemma, a concept recently examined in~\cite{Hui}.
	According to Hui \textit{et al.}~\cite{Hui}, for disjoint effective potentials, expanding the frequency domain Green's function as a sum of geometric series allows attributing successive echo pulses to individual terms in the expansion.
	In particular, the first term gives rise to a time-domain waveform identical to the original black hole, consistent with the causality requirement.
	From such an analysis, the echo modes and the instability of the fundamental mode become somehow obscure since none of the terms in Green's function implies any pole associated with the echo modes. 
	To recover the echo modes, we argued that one has to include an infinite number of terms in the summation of the geometric series.
	In this regard, echo is interpreted as a collective phenomenon that involves the entire spectrum of echo modes.
	To verify our statement and, in particular, ascertain that these echo modes are not merely artifacts, we explicitly extract the echo modes from the time-domain waveforms using Fourier analysis.
	The numerical findings confirm the presence of echo modes embedded within the waveforms, demonstrating their precise locations.
	In addition, we mention the difficulties encountered when applying the Prony method to the task.
	Although it has been firmly established as one of the most accurate and efficient methods to extract quasinormal frequencies from time-domain waveforms, the Prony method becomes somewhat restrictive when employed to waveforms with echoes.
	In particular, the obtained frequency values depend on the specific choice of the time interval of the analysis, and for the same time interval, time series with different resolutions offer different results.
	Also, as reported by Berti \textit{et al.}~\cite{BertiXfit}, it is worth noting that while the fundamental mode shows minimal uncertainty, significant discrepancies in echo modes are observed between analytic and numerical fitting approaches.
	In this regard, the echo modes are an intriguing problem and deserve further investigation owing to their immediate observational implications.
	
	
	\appendix
	
	\section{The derivation of Eq.~\eqref{asymptoticRegionFSimplified}}\lb{app1}
	
	In this Appendix, we give an account of the derivation of Eq.~\eqref{asymptoticRegionFSimplified} utilized in the main text.
	
	Since $L\gg 1$, the Wronskian Eq.~\eqref{Wrons2} can be simplified, and the equation that determines the QNM spectrum takes the form
	\begin{equation}
		\begin{array}{ll}
			2 i \beta \kappa \left[\omega+\tilde{\omega}-e^{2 i \sigma \tilde{\omega}}(\omega-\tilde{\omega})\right]
			~   _2F_1(1+\beta, \beta-i\omega/\kappa, 1-i\omega/\kappa; -e^{2\kappa L}) \\
			+\left[\omega^2 (1-e^{2i \sigma \tilde{\omega}}) +2 \omega \tilde{\omega} (1+e^{2i \sigma \tilde{\omega}}) +\tilde{\omega}^2 (1-e^{2i \sigma \tilde{\omega}})\right] ~ _2F_1(\beta, \beta-i\omega/\kappa, 1-i\omega/\kappa; -e^{2\kappa L})=0.
			\label{}
		\end{array}  
	\end{equation}
	We then can use the following identity for hypergeometric functions
	\bqn
	_2F_1(a, b,c; z)=(1-z)^{-a}~   _2F_1(a, c-b,c; \frac{z}{z-1}) 
	\eqn
	to obtain
	\begin{equation}
		\begin{array}{ll}
			2 i \beta \kappa \left[\omega+\tilde{\omega}-e^{2 i \sigma \tilde{\omega}}(\omega-\tilde{\omega})\right]
			(1 +e^{2\kappa L})^{-1}~   _2F_1(1+\beta, 1-\beta, 1-i\omega/\kappa; \frac{e^{2\kappa L}}{e^{2\kappa L}+1}) \\
			+\left[(\omega+\tilde{\omega})^2-e^{2 i \sigma \tilde{\omega}}(\omega-\tilde{\omega})^2\right] ~ _2F_1(\beta, \beta-i\omega/\kappa, 1-i\omega/\kappa;\frac{e^{2\kappa L}}{e^{2\kappa L}+1})=0,
			\label{}
		\end{array}  
	\end{equation}
	which can further be simplified in the large $L$ limit to get
	\begin{equation}
		\begin{array}{ll}
			2 i \beta \kappa \left[\omega+\tilde{\omega}-e^{2 i \sigma \tilde{\omega}}(\omega-\tilde{\omega})\right]
			e^{-2\kappa L}~   _2F_1(1+\beta, 1-\beta, 1-i\omega/\kappa; 1-e^{-2\kappa L}) \\
			+\left[(\omega+\tilde{\omega})^2-e^{2 i \sigma \tilde{\omega}}(\omega-\tilde{\omega})^2\right] ~ _2F_1(\beta, \beta-i\omega/\kappa, 1-i\omega/\kappa;1-e^{-2\kappa L})=0.
			\label{Eq: Wrons3}
		\end{array}  
	\end{equation}
	In the limit $L\gg 1$, we can expand the hypergeometric functions to obtain the following series
	\bqn
	\begin{array}{ll}
		_2F_1(1+\beta, 1-\beta, 1-i\frac{\omega}{\kappa};z )=-\frac{i \pi  \mbox{csch}(\frac{\pi \omega}{\kappa})\Gamma(1-i\frac{\omega}{\kappa})}{\Gamma(-\beta-i\frac{\omega}{\kappa}) \Gamma(\beta-i\frac{\omega}{\kappa})\Gamma(2+  i\frac{\omega}{\kappa})}+(1-z)^{-i\frac{\omega}{\kappa}-1}\frac{\omega}{\beta \kappa}\mbox{csch}(\frac{\pi \omega}{\kappa})\sin(\pi \beta)+\mathcal{O}[1-z]\\
		_2F_1(\beta, 1-\beta, 1-i\frac{\omega}{\kappa};z)= \frac{i \pi  \mbox{csch}(\frac{\pi \omega}{\kappa})\Gamma(1-i\frac{\omega}{\kappa})}{\Gamma(1-\beta-i\frac{\omega}{\kappa}) \Gamma(\beta-i\frac{\omega}{\kappa})\Gamma(1+  i\frac{\omega}{\kappa})}-i(1-z)^{-i\frac{\omega}{\kappa}}   \mbox{csch}(\frac{\pi \omega}{\kappa})\sin(\pi \beta)+\mathcal{O}[1-z],
		\label{Eq: Series}
	\end{array}
	\eqn
	where we have used the branch $0 \le \theta < 2\pi$ in the complex plane.  
	Combining Eqs.~(\ref{Eq: Wrons3}) and~(\ref{Eq: Series}) gives
	\bqn
	\begin{array}{ll}
		\frac{\pi \kappa \Gamma(1-i\frac{\omega}{\kappa})}{\Gamma(-\beta-i\frac{\omega}{\kappa}) \Gamma(\beta-i\frac{\omega}{\kappa})\Gamma(1+  i\frac{\omega}{\kappa})}  \left[ \frac{2 i \beta \kappa e^{-2 \kappa L}(\omega +\tilde{\omega}- e^{2 i \sigma \tilde{\omega}}(\omega -\tilde{\omega}))}{i \omega +\kappa}  
		-\frac{e^{2 i \sigma \tilde{\omega}}(\omega -\tilde{\omega})^2 
			-(\omega +\tilde{\omega})^2}{i \omega +\beta \kappa}\right]\\
		-\left(\omega^2-\tilde{\omega}^2\right) \sin (\pi  \beta ) \left(1-e^{2 i \sigma  \tilde{\omega}}\right) e^{2 i \omega L}=0.\label{Eqfff4}
	\end{array}
	\eqn
	
	By comparing Eq.~\eqref{Eqfff4} against the form Eq.~\eqref{Eq: fw}, we have
	\bqn
	\mathbf{f}(\omega)=\frac{\pi  \Gamma(1-i\frac{\omega}{\kappa})[(\omega +\tilde{\omega})^2 -e^{2 i \sigma \tilde{\omega}}(\omega-\tilde{\omega})^2]}{\left(\omega^2-\tilde{\omega}^2\right) \sin (\pi  \beta ) (1-e^{2 i \sigma \tilde{\omega}}) \Gamma(1-\beta-i\frac{\omega}{\kappa}) \Gamma(\beta-i\frac{\omega}{\kappa})\Gamma(1+  i\frac{\omega}{\kappa})} .
	\lb{LargeLLimit}
	\eqn
	We now look at the asymptotic region where $\omega \sim \omega_R \gg 1$.  
	In this region, we find
	\bqn
	\mathbf{f}(\omega) \approx -   \frac{2 i \omega^2 e^{\pi \omega / \kappa}}{ \tilde{\epsilon} \sin(\pi \beta)(1-e^{2 i \sigma \omega})}.
	\lb{asymptoticRegionF}
	\eqn
	where $\tilde{\epsilon}=V_\mathrm{PT}\left(L-\frac{\sigma}{2}\right)+\epsilon$.  
	Note that $|\omega_I|$ increases as $\omega_R$ increases.  
	For a relatively large $|\omega_I|$, one can further simplify the above expression as
	\begin{equation}
		\mathbf{f}(\omega) \approx  \frac{2 i \omega^2 e^{\pi \omega / \kappa}}{ \tilde{\epsilon} \sin(\pi \beta)e^{2 i \sigma \omega}}.
		\tag{\ref{asymptoticRegionFSimplified}}
	\end{equation}

\end{document}